\documentclass[lettersize,journal]{IEEEtran}
\usepackage{float}
\usepackage{amsmath,amsfonts}
\usepackage{algorithmic}
\usepackage{algorithm}
\usepackage{array}
\usepackage[font=normalsize,labelfont=sf,textfont=sf]{subcaption}
\usepackage{textcomp}
\usepackage{stfloats}
\usepackage{url}
\usepackage{verbatim}
\usepackage{graphicx}
\usepackage{cite}
\usepackage{booktabs}
\usepackage{amssymb}
\usepackage{tabularx}

\usepackage{tikz}

\usetikzlibrary{positioning, shapes, arrows}

\begin{document}

\title{
A Coupled V2G Equilibrium Model of Electric Vehicle and Power System 
Interactions}

\author{Jiaxin Hou, 
        Yujia Li,~\IEEEmembership{Member,~IEEE,}
        Jong-Shi Pang
\thanks{Jiaxin Hou and Jong-Shi Pang are with the Daniel J. Epstein Department of Industrial and Systems Engineering, University of Southern California, Los Angeles, California
90089-0193, U.S.A. (e-mail:jiaxinho@usc.edu; jongship@usc.edu).}
\thanks{Yujia Li is with Lawrence Berkeley National Laboratory (e-mail: yujiali@lbl.gov).}
\thanks{Jiaxin Hou and Jong-Shi Pang are supported by the U.S.\ Air Force Office of Scientific Research under grant FA9550-22-1-0045.}}



\maketitle

\begin{abstract}
Vehicle-to-grid (V2G) technology empowers electric vehicles (EVs) to act as mobile energy resources, providing critical support to power systems, especially under stressed conditions. To understand the economic mechanism driving V2G participation and its benefits to power grid, this paper proposes a multi-player coupled equilibrium framework that models the bidirectional interactions between power grid operations and EV routing, incorporating charging and discharging choice in a preprocessed feasible path generation procedure. Energy prices are endogenously determined by market clearance conditions. We formulate the overall problem as a Variational Inequality that unite the decision-making of Distribution System Operator, Charging Network Operator, Load Serving Entities, and EV drivers.
Numerical studies validate the framework under two stress scenarios: increased household load and power line outages. 
Results show that when EVs are incentivized by reduced generalized path costs, V2G is particularly effective in eliminating load shedding and reducing distribution locational marginal electricity prices.
On the transportation side, V2G can 
lead to divergence in EV behavior between normal and scarcity conditions, and alter route choices yet improve overall trip economic.



\end{abstract}


\vspace{-0.1in}

\section*{Nomenclature}

\subsection*{Sets and Indices}
\subsubsection*{Power Network}
\begin{IEEEdescription}[\IEEEsetlabelwidth{$P_{loss, max}$}]
  \item[$d, i, (i,j)$] Indices for loads, buses, and lines.
  \item[$i_{\rm tso}$] Index for node connected to the higher-level transmission system operator (TSO).
  \item[$g, f$] Indices for distributed generators (DGs) and 
  load service entities (LSEs).
  \item[$\Omega^B, \Omega^L, \mathcal{F}$] Sets of buses, distribution lines and LSEs.
  \item[$\Omega^G_f$] Sets of buses where LSE $f$ operates generators.
  \item[$\Omega^{B}_{-{\rm tso}}$] $\triangleq \Omega^{B}\setminus \{i_{\rm tso}\}$ 
\end{IEEEdescription}

\vspace{0.1in}
\subsubsection*{Transportation Network}
\begin{IEEEdescription}[\IEEEsetlabelwidth{$P_{loss, max}$}]
\item [$n,nm/a$] Indices for nodes $n$, arcs $(n,m)$, in this paper, we interchangeably use $(m,n)$ and $a$ for arcs
\item [$n', n''$] Dummy charging and discharging node for node $n$ with a fast charging stations (FCS).
\item [$q,rs$] Indices for paths, origin-destination (OD) pairs.
\item [$\mathcal{Q}_{rs},\mathcal{Q}_{rs}'$] Set of paths for OD pair with origin $r$ and destination $s$ in the original network ($\mathcal{Q}_{rs})$ and the extended network ($\mathcal{Q}_{rs}'$).
\item [$\mathcal{S}$] Set of all OD pairs.
\item [$\mathcal{N}$] Set of transportation nodes in the original network.
\item [$\mathcal{N}^{\rm fcs}$] Set of all nodes with fast charging
stations (FCSs) in the original network.
\item [$\mathcal{N}^q$] Set of nodes on path $q$ in the original network.
\item [$\mathcal{N}^{q}_{\pm}$] Set of nodes on path $q$ in the expanded network where the vehicle charges (+) or discharges (-).
\item [$\mathcal{N}^{q}_{\rm act}$] $\triangleq \mathcal{N}^{q}_+ \cup \mathcal{N}^{q}_-$ .
\item [$\mathcal{A}$] Set of all arcs in the original network. 
\item [$\Omega^{\rm fcs}$] Set of FCSs managed by CNO.
\item [$\Omega_i^{\rm fcs}$] Set of FCSs connected to power bus $i$.
\end{IEEEdescription}

\vspace{-0.1in}
\subsection*{Parameters}
\subsubsection*{CNO and Power Network}
\begin{IEEEdescription}[\IEEEsetlabelwidth{$P_{loss, max}$}]
  \item [$\underline{P}_{n}^{\rm fcs}, \overline{P}_{n}^{\rm fcs}$] Lower and upper bounds on net power exchange at FCS $n$.
  \item [$r_{ij},\,x_{ij}$] Resistance and reactance of line $ij$.
  \item [$S^{\max}_{ij}$] Capacity rating of line $ij$.
  \item [$\underline{U}^2, \overline{U}^2$] Lower and upper bounds on squared voltage magnitude at bus $i$.
  \item [$\underline{P}_{\rm tso}, \overline{P}_{\rm tso}$] Lower and upper bounds on active power imported from TSO.
  \item [$\underline{Q}_{\rm tso}, \overline{Q}_{\rm tso}$] Lower and upper bounds on reactive power imported from TSO.
  \item [$q_i^{\rm load}$] Fixed reactive load at each bus $i$.
  \item [$a_i, b_i$] Linear and fixed sales quantity coefficients at bus $i$. $a_i<0, b_i>0$.
  \item [$d_{f,i}, e_{f,i}$] Linear and fixed cost coefficients of dispatchable generator located at bus $i$ and operated by LSE $f$. $d_{f,i}>0, e_{f,i}>0$.
  \item [$\underline{p}^{\mathrm{gen}}_{f,i}, \overline{p}^{\mathrm{gen}}_{f,i}$] Minimum and maximum generation capacity of LSE $f$ at bus $i$.
  \item [$p_i^{\rm demand}$] Given portion of household demand at but $i$.
  \item [$\rho$] Load shedding penalty.
\end{IEEEdescription}

\vspace{0.1in}
\subsubsection*{Transportation Network}
\begin{IEEEdescription}[\IEEEsetlabelwidth{$P_{loss, max}$}]
\item [$d_{nm}$] Traveling distance on arc $(n,m)$.
\item [$\beta_{nm}$] Energy consumption rate for the arc $(n,m)$.
\item [$B_{\max}$] The EV battery capacity (in kWh). 
\item [$B_{\min}$] Post-discharge energy level.
\item [$R^{\rm anx}$] Range anxiety rate of EVs (e.g., 20\%).
\item [$E^{q,\pm}_{n'}$] Charging/discharging energy at $n'$(+)/$n''(-)$ on path $q$ (determined, in kWh/flow).
\item [$w^{x}$] Weight factor converting time to monetary cost for vehicle $x$, where $x\in\{\rm EV, \rm FV\}$.
\item [$t^0_{nm},c_{nm}$] Free-flow travel time, capacity for arc $(n,m)$. 
\item [$c_{n}$] Vehicular capacity at FSC at node $n$. 
\item [$\Delta t$] Study time period.
\item [$D^{x}_{rs}$]  Total travel demand of vehicle type $x$ for OD pair $rs$, where $x\in\{\rm EV, \rm FV\}$.
\item [$t^0_n,\widehat{t}_n$] Base waiting time, congestion parameter at $n$.\\[3pt]
\item [$\overline P^{\rm evcs}$] Nominal power capacity of charging piles in FCSs. 
\item [$c_{\rm deg}$] Battery degradation per kWh.
\end{IEEEdescription}
%
%
%
%
%
%
\subsection*{Decision Variables}
\subsubsection*{CNO and Power Network}
\begin{IEEEdescription}[\IEEEsetlabelwidth{$P_{loss, max}$}]
  \item [$p_{\rm cno,n}^{\rm fcs,\pm}$] Power quantity sells to Electric Vehicles (EVs), which is the same as the quantity purchase from (+) and sells to (-) LSEs
  \item [$\widehat{P}_{i}^{\rm dso}, \widehat{P}_{i}^{\rm tso}$] Power injection at bus $i$ from non-TSO $(\widehat{P}_{i}^{\rm dso})$ and TSO sources $(\widehat{P}_{i}^{\rm tso})$, scheduled by the DSO.
  \item [$U^{sqr}_{i }$] Squared voltage magnitude at bus $i$.
  \item [$p^{line}_{ij },\,q^{line}_{ij }$] Active/reactive power flow on line $ij$.
  \item [$p^{\mathrm{gen}}_{f,i},p^{\mathrm{sell}}_{f,i}$] LSE $f$ active power generation and power households sales at bus $i$.
  \item [$\phi_{f,i}^{\rm lse, \pm}$] LSE $f$'s sale (+) and purchase (-) quantities with the CNO.
    \item [$\phi_{f,i}^{\rm lse}$] LSE $f$'s purchase quantity from TSO.
    \item [$LS$] Load shedding amount.
\end{IEEEdescription}

\subsubsection*{Transportation Network}
\begin{IEEEdescription}[\IEEEsetlabelwidth{$P_{loss, max}$}]
\item [$f_{rs}^{q,x}$]  Traffic flow of EVs and FVs on path $q$ for OD pair $rs$, where $x \in \{\rm EV, \rm FV\}$.
\item [$C_{rs}^{x}$]  Equilibrium cost of EVs and FVs for OD pair $rs$, where $x \in \{\rm EV, \rm FV\}$. 
\end{IEEEdescription}
\vspace{-0.1in}
\subsection*{Induced Variables}
\subsubsection*{Power System}
\begin{IEEEdescription}[\IEEEsetlabelwidth{$P_{loss, max}$}]
\item [$P^{\mathrm{sell}}_{i}$] Total power sold to household at bus $i$.
\end{IEEEdescription}

\subsubsection*{Transportation System}
\begin{IEEEdescription}[\IEEEsetlabelwidth{$P_{loss, max}$}]
\item [$x_{nm}$] Traffic flow through arc $(n,m)$. 
\item [$x_n^{\pm}$] Traffic flow through the dummy charging (+) and discharging (-) node, respectively.
\item [$x_n$] Traffic flow through the FCS at node $n$. 
\item [$t_{nm}$] Travel time on arc $(n,m)$.
\item [$W_n$] Waiting time at FCS node $n$.
\item [$\mathrm{EF}^{\pm}_{n}$] Total charging (+),  discharging (-) energy at FCS $n$, respectively. $\mathrm{EF}_{n} = \mathrm{EF}^{+}_{n} + \mathrm{EF}^{-}_{n}$
\end{IEEEdescription}

\vspace{-0.1in}
\subsection*{Price Signals}
\begin{IEEEdescription}[\IEEEsetlabelwidth{$P_{loss, max}$}]
\item [$w_{x, i}$] Locational marginal price at power bus $i$ from $x$ perspective, where $x \in \{{\rm dso}, {\rm lse}\}$.
\item [$\alpha_{x,n}^{\pm}$] Charging (+)  and discharging (-) prices at FCS $n$ from $x$ perspective, where $x \in \{{\rm cno}, {\rm UE}\}$.
\item [$M_i^{x, \pm}$] CNO-LSE purchasing (+) and selling (-) prices at bus $i$ from $x =\rm cno$ perspective and the opposite for $x = \rm lse$.
\item [$m_i^x$] TSO-LSE Transaction price at bus $i$ from $x$ perspective where $ x\in \{\rm tso, \rm lse\}$.
\end{IEEEdescription}
%
%
%
%
%
%
%
\subsection*{Shared Variables in Market-Clearing Conditions}
\begin{IEEEdescription}[\IEEEsetlabelwidth{$P_{loss, max}$}]
\item [$w_{i}$] Locational marginal price at power bus $i$. 
\item [$\alpha_n^{\pm}$] Charging (+)/discharging (-) prices at FCS $n$.
\item [$M_i^{\pm}$] Purchasing (+) and selling (-) price at bus $i$ between CNO and LSEs.
\item [$m_i$] Selling price at bus $i$ between TSO and LSEs.
\end{IEEEdescription}
\vspace{-0.1in}
\subsection*{Functions}
\begin{IEEEdescription}[\IEEEsetlabelwidth{$P_{loss, max}$}]
\item [$C_{rs}^{q,x}$]  Path cost of vehicle type $x$ for OD pair $rs$, where $x\in\{\rm EV, \rm FV\}$.
\item [$C_{f,i}$] Generation cost function of LSE $f$ at bus $i$.
\item [$R_i$] Inverse demand function (unit price) at bus $i$.
\end{IEEEdescription}
\vspace{0.2in}

\section{Introduction}\label{sec:intro}

\IEEEPARstart{T}{he} transition toward a sustainable energy future has rapidly accelerated the electrification of the transportation sector. Global electric vehicle (EV) sales exceeded 17 million in 2024, accounting for more than 20\% of new cars sold worldwide, with projections indicating this market share will surpass 40\% by 2030 under existing policy trajectories~\cite{iea2025}. 
The mass adoption of EVs has introduced an intricate coupling between transportation networks and electric power grids. As EV penetration scales, the spatio-temporal distribution of the charging demand threatens to induce operational stress on local distribution systems, such as cable overloading~\cite{lei2024}, voltage unbalance~\cite{yong2015review}, and power losses~\cite{clement2010, fernandez2010assessment}. However, this growing interdependence also presents a transformative opportunity through the emergence of vehicle-to-grid (V2G) technology. Through bidirectional power flow, EVs can function as both loads and mobile energy resources capable of supplying electricity back to the grid. This capability allows EVs to
contribute to grid stability and flexibility, reduce peak load, and alleviate grid congestion, thereby enhancing overall power system resilience~\cite{alamgir2025, eltohamy2025,
rana2024, pang2025,zhan2025,xiao2025}.

\noindent Most existing works, however, adopt optimization- or control-based models that do not explicitly capture the EV routing, charging decisions, and distribution-grid operation within an interacting framework. In practice, these decisions emerge from
the strategic behavior of interacting agents linked through prices, physical network constraints, and market-clearing
conditions. This renders the problem fundamentally one of coupled equilibrium, making it essential to adopt an equilibrium framework that explicitly captures interactions among actors across both the transportation and power systems. The interactions inherently run in two directions: charging and discharging prices directly influence traveler behavior, while the resulting traffic and charging patterns concurrently shape grid loading and operating feasibility. This perspective becomes especially valuable under stressed operating conditions, where fast charging stations (FCSs) actively participate in the equilibrium outcome not merely as loads, but as critical sites of V2G discharging to support grid resilience.

\noindent Current equilibrium models for coupled transportation and power systems primarily focus on traffic assignment based on Wardrop's user equilibrium and power system operations. Earlier works, such as Wei et al.~\cite{wei2017network}, have laid the foundation by proposing a united model that integrates transportation and power distribution systems, capturing the interdependencies between EV behaviors and electricity prices. More recent literature includes Stackelberg-Nash frameworks where the Power Distribution Network acts as the leader, determining charging prices, while drivers respond with routing adjustments~\cite{li2025network}, as well as dynamic models that consider realistic user and traffic behaviors~\cite{song2025dynamic}.
While these models offer insights into interactions between EV routing and power system operation, there remains an opportunity to incorporate V2G capabilities.
Moreover, existing V2G studies have not explicitly modeled the economic incentives that drive EV participation. This limits the ability to assess the realistic adoption and effectiveness of V2G in practice. To address this, our proposed framework explicitly models EV incentives alongside the interactions between traffic routing, charging/discharging choices, and multi-agent decision-making, whereby electricity prices are determined endogenously through market clearing.
As such, our framework provides infrastructure planners with actionable insights to accelerate the strategic deployment of V2G-enabled charging stations and ensure their economic viability in a competitive market.

\noindent As the term V2G suggests, the proposed (noncooperative)
game-theoretic framework consists of two distinct components with
multiple self-interested players. On one hand, the transportation subsystem is modeled as a traffic equilibrium problem based on Wardrop’s principle of route choice over a transportation network with fixed demand to be met. On the other hand, the power subsystem comprises a regional distribution network operated by a Distribution System Operator (DSO), multiple Load Serving Entities (LSEs), and a Transmission System Operator (TSO) which supplies electricity to the local grid through the wholesale market. The DSO acts as the primary wholesale buyer, procuring power from the TSO to resell to LSEs. Beyond this transaction, the DSO serves exclusively as a local network distributor, facilitating physical power delivery and managing distribution-level constraints. These two subsystems are coupled via the Charging Network Operators (CNO), which manages the charging infrastructure FCSs and engages in energy transactions with both LSEs and EV users. In this context, EV charging (grid-to-vehicle) and discharging (V2G) are realized as power withdrawals and injections, respectively, at the distribution nodes. 
For tractability, we assume a single CNO, DSO, and TSO, while allowing multiple LSEs to compete strategically.

\noindent The main contributions of this paper are threefold. First, we propose a coupled equilibrium framework for transportation and power systems with V2G integration. Second, we model EV charging and discharging behavior within a user equilibrium setting, explicitly capturing incentive-driven participation. Third, through numerical analysis of extreme operating scenarios, we quantify the system-level benefits of V2G and provide actionable insights for infrastructure planners regarding the economic and operational viability of deploying V2G-enabled FCSs.

\noindent The remainder of this paper is organized as follows. Section~\ref{sec:CNO} introduces the CNO model, followed by the transportation network equilibrium formulation in Section~\ref{sec:traffic}. Section~\ref{sec:DSO} presents the DSO model, and Section~\ref{sec:LSEs} describes the LSEs’ decision-making framework. The market-clearing conditions that couple all system components are given in Section~\ref{sec:mc}, while Section~\ref{sec:overall problem} presents the overall equilibrium formulation as a variational inequality. Numerical experiments and results are discussed in Section~\ref{sec:numerical}. Finally, Section~\ref{sec:conclusion} concludes the paper.
\section{Mathematical Formulations}
\noindent This section details the mathematical formulation of the proposed game-theoretic framework. We characterize the decision-making process of each player.

\subsection{The Charging Network Operator (CNO) Module} \label{sec:CNO}
\noindent The CNO manages a set of FCSs where EVs can charge and discharge. Anticipating the charging(+)/discharging(-) prices $\{ \alpha_{{\rm cno};n}^{\pm} \}_{ n \in \Omega^{\rm fcs}}$ and the locational electricity prices 
$\{ M_i^{\rm cno,\pm} \}_{i \in \Omega^B}$, the CNO decides how much power $p_{{\rm cno};n}^{{\rm fcs};\pm}$ to sell to and buy from EVs at each FCS $n \in \Omega^{\rm fcs}$ to maximize net profit; thus the CNO's decision is a maximization problem:
\small{\begin{subequations}\label{eq:CNO_basic}
\begin{align}
&\max_{\substack{p_{{\rm cno},n}^{\rm fcs,\pm}}\geq0}  
\sum_{n\in \Omega^{\rm fcs}}
\Big\{\underbrace{
(\gamma_n^+ + \alpha_{{\rm cno},n}^{+})\,p_{{\rm cno},n}^{\rm fcs,+}}_{\text{Revenue from EV}}
\;-\;\underbrace{(\alpha_{{\rm cno},n}^{-}-\gamma_n^-)\,p_{{\rm cno},n}^{\rm fcs,-}}_{\text{Payments to EV}}  \Big\}
\nonumber\\
&\,-\,
\sum_{i\in\Omega^{B}_{-\rm tso}}
\sum_{
n\in \Omega_i^{\rm fcs}}
\Big\{
\underbrace{M_{i}^{\rm cno,+}p_{{\rm cno},n}^{\rm fcs,+}}_{\text{Payment to LSE}}
-\underbrace{M_{i}^{\rm cno,-}p_{{\rm cno},n}^{\rm fcs,-}}_{\text{Revenue from LSE}}\Big\}
\label{CNO_OBJ}
\\
&\text{s.t.}\quad
\underline{P}_{n}^{\rm fcs}
{\le} 
p_{\rm cno, n}^{\rm fcs,+} - p_{\rm cno,n}^{\rm fcs,-}
{\le}
\overline{P}_{n}^{\rm fcs}, 
\quad n\in\Omega^{\rm fcs}
\label{CNO_capacity}
\end{align}
\end{subequations}}
\subsection{Traffic Network Equilibrium} \label{sec:traffic}
\noindent The traffic module characterizes the routing behavior of both EVs and FVs within a user equilibrium (UE) framework. For EVs, routing decisions are coupled with charging and discharging decisions. To maintain tractability, the network is extended through a pre-processing procedure that embeds charging and discharging decisions into the network topology.

\subsubsection{Expanded network}

\begin{figure}[!t]
\centering
\includegraphics[width=1\linewidth]{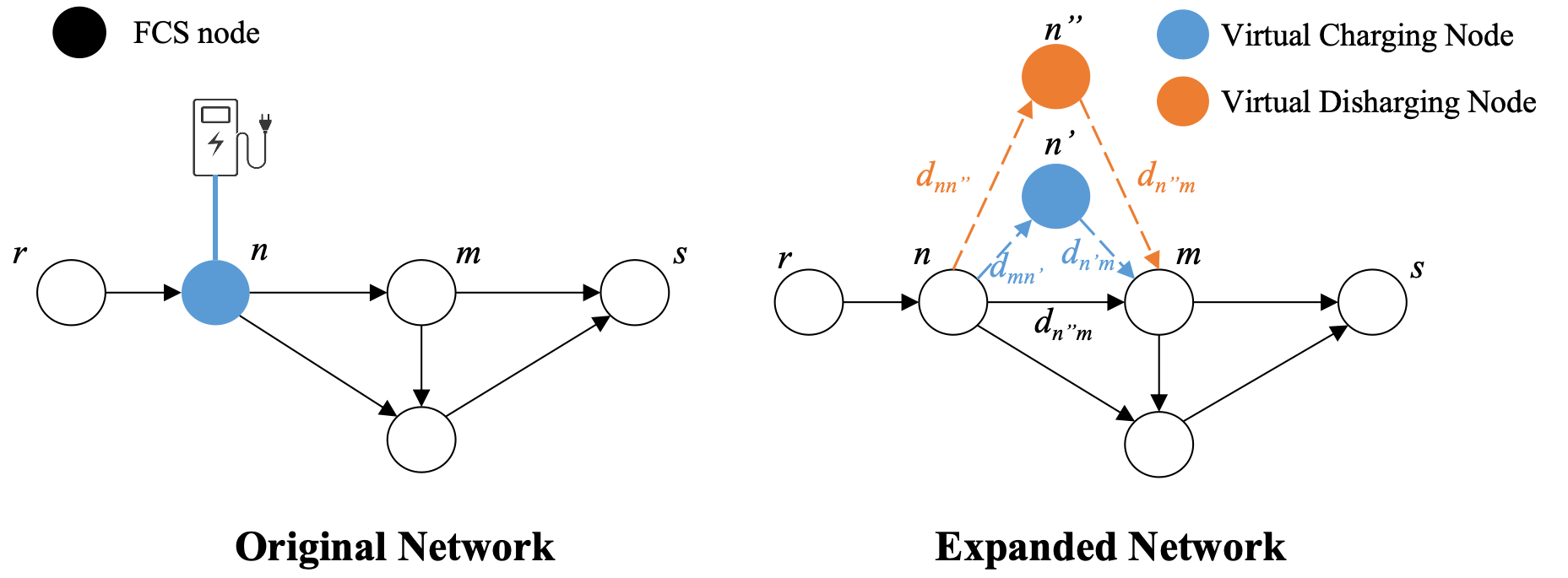}
\caption{Original and Expanded Transportation Networks}
\label{expanded_network}
\end{figure}

\noindent  Figure~\ref{expanded_network} illustrates the original and expanded transportation networks. In the original network (left), EVs travel along standard transportation nodes, with some nodes equipped with FCSs where EVs can charge or discharge energy. In the expanded network (right), virtual nodes explicitly represent charging and discharging decisions. Specifically, each FCS node \( n \) has two virtual nodes: a virtual charging node \( n' \) (blue) and a virtual discharging node \( n'' \) (orange). Corresponding virtual arcs connect these nodes back to the physical network. Distances along these virtual arcs satisfy \( d_{nn'} = d_{nn''} = 0 \), indicating no travel distance for initiating charging or discharging at the station node itself. However, virtual arcs from these nodes back to the physical network maintain the original arc distances, i.e., $d_{n'm} = d_{n''m} = d_{nm}$, ensuring accurate modeling of travel distances. This representation explicitly tracks EV energy states and charging decisions along their routes. In particular, for the dummy arcs, we have the following conditions:
\small{\begin{equation*}
\begin{aligned}
d_{nn'} &= d_{nn''} = 0, \quad 
\beta_{n'n} = \beta_{n''n} = 0, \\
&\forall 
\, 
n\in\mathcal{N}^{\rm fcs},\;
n'\in \mathcal{N}^{+}_n,\; n''\in\mathcal{N}^{-}_n;\\[3pt]
d_{n'm} &= d_{n''m} = d_{nm}, \quad 
\beta_{n'm} = \beta_{n''m} = \beta_{nm}, \\
&\forall \, 
n\in\mathcal{N}^{\rm fcs}\;
n'\in \mathcal{N}^{+}_n,\; n''\in\mathcal{N}^{-}_n.
\end{aligned}
\end{equation*}}

\subsubsection{Pre-processing of Feasible Paths}
\label{sec:pathgen}

To eliminate the path-dependent charging amount $\{E_{n'}^{q',+}\}$ and discharging amount 
$\{E_{n''}^{q',-}\}$ variables in the traffic assignment stage, we adopt
the following behavioral and physical assumptions.

\begin{itemize}
  \item[\textbf{A1}] \textbf{Limited station visits.}  
        An EV may interact with FCSs at most twice during a single trip. Empirical data suggest that
        range-anxious drivers are wheeling to add one charging stop, and occasionally a second stop if energy discharge is allowed for
        arbitrage. More than two detours would rarely be chosen for convenience reasons.

  \item[\textbf{A2}] \textbf{Trips start with a full battery.}  
        Every EV departs its origin with the battery at maximum capacity
        \(B_{\max}\).  
        This assumption is consistent with public-charging studies that drivers typically begin trips with high state of charge (e.g., \cite{NREL2024ChargingSurvey}). It eliminates the need for path-dependent initial energy variables and ensures that charging decisions arise only when the state of charge has decreased
        predictably along the chosen route.
        
  \item[\textbf{A3}] \textbf{Greedy discharge behaviour.}  
       When an EV chooses to discharge at a FCS, it supplies energy up to a prescribed minimum state of charge, removing the possibility of multiple discharging events. Specifically, the discharged amount is such that the post-discharge energy level equals \(B_{\min}\), where \(B_{\min} > \underline{E} = R^{\rm anx}B^{\rm max}\), while ensuring sufficient charge to reach the destination or the next charging stop. 
       
     \item[\textbf{A4}] \textbf{Greedy charge behaviour.}  
        Conversely, when an EV selects to charge at an FCS, the driver top-ups
        to the highest admissible battery level. Hence the post-charging
        energy state is forced to the upper bound
        \(B_{\max}\), removing the continuous
        decision on partial charging.


\end{itemize}

\noindent Together, assumptions~\textbf{A1}–\textbf{A4} ensure that for any OD pair $(r,s)$ the energy balance along a
path $q'$ can be evaluated \emph{before} solving the equilibrium, turning
$E_{n'}^{q',+}$ and $E_{n''}^{q',-}$ into fixed path attributes rather than
variables. In what follows, we identify the set of energy-feasible paths in the expanded network for each OD pair and at the same time, show how the charged and discharged amounts at the FCSs can be pre-computed for every feasible path.

\subsubsection{Path pattern and feasibility conditions}\label{sec:pathpattern}

Under assumptions~\textbf{A1}--\textbf{A4}, we consider only \emph{energy-feasible} paths $q\in\mathcal{Q}'_{rs}$. Feasibility requires sufficient energy to (i) reach the first FCS, (ii) travel between FCSs, and (iii) reach the destination. 
Let path $q$ consist of an ordered sequence of nodes $(n_0, n_1, \dots, n_K, n_{K+1})$, where $n_0$ is the origin, $n_{K+1}$ is the destination, and $n_{k_1}, n_{k_2}$ be two distinguish intermediate charging/discharging nodes with $0\leq k_1<k_2\leq K+1$. Limiting to at most two stops yields the following feasibility condition, which are summarized in Table~\ref{tab:virtual_nodes}, 
\begin{align}\label{eq:feas_cond}
E_{n_{k_1}}^{\rm dep} - \sum_{a \in q(n_{k_1}, n_{k_2})} \beta_a d_a \ge E_{n_{k_2}}^{\rm arr\_min},
\end{align}
where the departure energy $E_{n_{k_1}}^{\rm dep}$ and minimum required arrival energy $E_{n_{k_1}}^{\rm arr\_min}$ are defined based on the node types:
\begin{align*}
E_{n_{k_1}}^{\rm dep} &=
\begin{cases}
B_{\max}, & \text{if } k_1=0 \text{ or } n_{k_1} \in \mathcal{N}^q_+ \\
B_{\min}, & \text{if } n_{k_1} \in \mathcal{N}^q_-
\end{cases} \\
E_{n_{k_2}}^{\rm arr\_min} &=
\begin{cases}
B_{\min}, & \text{if } n_{k_2} \in \mathcal{N}^q_- \\
\underline{E}, & \text{if } k_2 = K+1 \text{ or } n_{k_2} \in \mathcal{N}^q_+
\end{cases}
\end{align*}

\begin{table}[!t]
\caption{Feasible Path Patterns (with $(k_1,k_2)$)}
\label{tab:virtual_nodes}
\footnotesize
\centering
\setlength{\tabcolsep}{3pt}
\renewcommand{\arraystretch}{0.95}
\begin{tabular}{@{}lccc@{}}
\toprule
\textbf{Pattern} & \textbf{Virtual Nodes} & $(k_1,k_2)$ & \textbf{Comment} \\
\midrule
0 &  None & $(0,K\!+\!1)$ & No action \\[0.05in]
1-1 & $n_1'$ & $(n_1',K\!+\!1)$& Charge once\\[0.05in]
1-2 & $n_1',n_2'$ &  $((n_1',n_2'))$ & Charge twice\\[0.05in]
2 & $n_1''$ & $(n_1'',K\!+\!1)$ & Discharge once \\[0.05in]
3 & $n_1',n_2''$ & $(n_1',n_2'')$ & Charge $\rightarrow$ discharge \\[0.05in]
4 & $n_1'',n_2'$ & $(n_1'',n_2')$ & Discharge $\rightarrow$ charge \\
\bottomrule
\end{tabular}
\end{table}

\noindent\emph{Path costs} include travel time, waiting time, charging/discharging time, and expenses(+)/revenues(-) at anticipated prices $\{\alpha_{\rm UE;n}^{\pm}\}$. 
\begin{align*}
C_{rs}^{q;{\rm EV}} =
&\, w^{\rm EV} \left[
\sum_{a \in q} t_a
+ \sum_{n \in \mathcal{N}^q_{\rm act}} W_n
+  \frac{\displaystyle\sum_{n \in \mathcal{N}^q_+}E_n^{q,+}
+\displaystyle\sum_{n \in \mathcal{N}^q_-} E_n^{q,-}}
{\overline{P}^{\rm EVCS}}
\right] \\
+&\sum_{n \in \mathcal{N}^q_+} E_n^{q,+} (\alpha_{{\rm UE};n}^{+} + c_{\rm deg})
- \sum_{n \in \mathcal{N}^q_-} E_n^{q,-} (\alpha_{{\rm UE};n}^{-} - c_{\rm deg})\\
C_{rs}^{q;{\rm FV}} =
&\, w^{\rm FV}
\sum_{a \in q} t_a, \quad \text{if } q \in \mathcal{Q}_{rs}.
\end{align*}

\subsubsection{Path-generation procedure}
Algorithm~\ref{alg:pathgen} constructs the set of feasible paths by augmenting base routes with charging and discharging decisions. Feasibility is verified via a single forward traversal of each path, where the SoC is updated aligned with assumptions~\textbf{A1}–\textbf{A4}. A path is retained if it satisfies the feasibility conditions~\eqref{eq:feas_cond}.

\small
\begin{algorithm}[!t]
\caption{\textsc{Generate Paths} for O--D pair \((r,s)\)}
\label{alg:pathgen}
\begin{algorithmic}[1]
\STATE \textbf{Input:} base network \((\mathcal{N},\mathcal{A})\), FCS set \(\mathcal{N}^{\rm fcs}\)
\STATE \textbf{Output:} expanded path set \(\mathcal{Q}_{rs}'\)
\STATE \(\mathcal{Q}_{rs}' \gets \emptyset\)
\STATE \(\mathcal{B} \gets\) the \(K\)-shortest simple paths from \(r\) to \(s\)

\FOR{\(p \in \mathcal{B}\)}
    \IF{\textsc{Feasible}(p)}
        \STATE \(\mathcal{Q}_{rs}' \gets \mathcal{Q}_{rs}' \cup \{p\}\)
    \ENDIF

    \FOR{each \(n \in \mathcal{N}^{\rm fcs}\) on \(p\)}
        \FOR{each \(\tau \in \{n',\, n''\}\)}
            \STATE \(q \gets \textsc{Insert}(p,\tau)\)
            \IF{\textsc{Feasible}(q)}
                \STATE \(\mathcal{Q}_{rs}' \gets \mathcal{Q}_{rs}' \cup \{q\}\)
            \ENDIF
        \ENDFOR
    \ENDFOR

    \FOR{each ordered pair of distinct FCSs \((n_1,n_2)\) on \(p\)}
        \FOR{each \((\tau_1,\tau_2) \in \{(n'_1,n'_2),\,(n''_1,n'_2),\,(n'_1,n''_2)\}\)}
            \STATE \(q \gets \textsc{Insert}(p,\tau_1,\tau_2)\)
            \IF{\textsc{Feasible}(q)}
                \STATE \(\mathcal{Q}_{rs}' \gets \mathcal{Q}_{rs}' \cup \{q\}\)
            \ENDIF
        \ENDFOR
    \ENDFOR
\ENDFOR

\STATE \textbf{return} \(\mathcal{Q}_{rs}'\)
\end{algorithmic}
\end{algorithm}

\subsubsection{User Equilibrium for Route Choice}
\noindent Two user populations are considered: EVs and fuel-vehicles (FVs). The traffic equilibrium determines the path flows \(\{f_{rs}^{q',\rm EV},\, f_{rs}^{q,\rm FV}\}\) for each OD pair,
\begin{subequations}
\begin{align}
    0 \leq C_{rs}^{\rm EV}\perp& \; \sum_{q \in \mathcal{Q}_{rs}'} f_{rs}^{q,\rm EV} -  D^{\rm EV}_{rs}\geq 0, \quad
     \forall rs \in \mathcal{S}, 
    \label{eq:total_flow}
    \\[0.1in]
    0 \leq C_{rs}^{\rm FV}\perp& \; \sum_{q \in \mathcal{Q}_{rs}} f_{rs}^{q,\rm FV} - D^{\rm FV}_{rs} \geq 0, \quad
     \forall rs \in \mathcal{S}, 
    \label{fv:total_flow}
    \\
    0 \leq f_{rs}^{q,\rm EV} \perp& \;C_{rs}^{q,\rm EV} - C^{\rm EV}_{rs} \geq 0, \;
     \forall \, rs \in \mathcal{S},\, q \in \mathcal{Q}_{rs}', 
    \label{eq:user_equilibrium}
    \\[0.1in]
    0 \leq f_{rs}^{q,\rm FV} \perp& \;C_{rs}^{q,\rm FV} - C^{\rm FV}_{rs} \geq 0, \;
     \forall \, rs \in \mathcal{S},\, q \in \mathcal{Q}_{rs}.
    \label{fv:user_equilibrium}
\end{align}
\end{subequations}
along with the induced arc flows \(\{x_{nm}\}\) and FCS visit flows \(\{x_{n'}^+, x_{n''}^-\}\). Travel times on network arcs follow the Bureau of Public Roads (BPR) function and a simplified waiting-time model is adopted, where queuing occurs only when the total flow within a given time period exceeds the number of charging piles, and the waiting time is modeled as a linear function of the excess flow.
\begin{subequations}
\begin{align}
    t_{nm} &= t^0_{nm}\Bigg(1 + 0.15\Big(\frac{x_{nm}}{c_{nm}}\Big)^4\Bigg), \;
     \forall\; (n,m) \in \mathcal{A}, 
    \label{eq:arc_travel_time}
    \\
    t_{n} &= t^0_{n}  + \widehat{t}_{n} \max(x_{n}\Delta t - c_{n}, 0), \;
    \forall\; n \in \mathcal{N}^{\rm fcs},
    \label{eq:fcs_wait}
\end{align}
\end{subequations}
where arc flows are given by
\small{\begin{align}
x_{nm} = &\sum_{rs\in \mathcal{S}}
\Big\{
\sum_{q\in \mathcal{Q}_{rs}'} 
f^{q,\rm EV}_{rs}
\Big(
\sum_{n'\in\mathcal{N}^{q}_+} \delta^{q}_{n'm}
+ \sum_{n''\in\mathcal{N}^{q}_-} \delta^{q}_{n''m}
\Big)
\notag\\
+
&
\sum_{q \in \mathcal{Q}_{rs}} 
f_{rs}^{q,\rm FV}\,\delta_{nm}^q 
\Big\},
\quad \forall \; (n,m)\in \mathcal{A},
\label{eq:arc_flow}
\end{align}}
and the charging and discharging flows at FCS nodes satisfy
\begin{subequations}
\begin{align}
x_n^+ &= \sum_{rs\in \mathcal{S}}\sum_{q \in \mathcal{Q}_{rs}'} 
\sum_{n'\in\mathcal{N}^{q}}
f^{q,\rm EV}_{rs}\gamma_{n'}^{q,+},\;
\forall n\in \mathcal{N}^{\rm fcs}, 
\label{eq:charging_flow}
\\
x_n^- &= \sum_{rs\in \mathcal{S}}\sum_{q \in \mathcal{Q}_{rs}'}
\sum_{n''\in\mathcal{N}^{q}}
f^{q,\rm EV}_{rs}\gamma_{n''}^{q,-},\;
\forall n\in \mathcal{N}^{\rm fcs}, 
\label{eq:discharging_flow}
\\
x_n &= x_n^+ + x_n^-,\;
\forall n\in \mathcal{N}^{\rm fcs}.
\label{eq:node_flow_balance}
\end{align}
\end{subequations}
\noindent For computational ease, we use the following smoothed approximation of the waiting-time function $t_n$~\eqref{eq:fcs_wait}, 
\begin{equation}\label{eq:fcs_wait_smooth}
{W}_n
=
t_n^0
+
\frac{1}{2}\widehat{t}_n
\begin{cases}
0 & x_n\Delta t - c_n \le 0, \\[6pt]
(x_n\Delta t - c_n)^2 & 0 < x_n\Delta t - c_n \le \varepsilon, \\[6pt]
2\varepsilon (x_n\Delta t - c_n) - \varepsilon^2
& x_n\Delta t - c_n > \varepsilon.
\end{cases}
\end{equation}
With $\varepsilon = \frac{1}{\Delta t}$, the linear part of 
${W}_n$ has the same slope as the original $t_n$ 
whenever $x_n \Delta t - c_n > \varepsilon$.
\subsubsection{From EV Traffic Flows to Energy Flows}
\noindent Equations~\eqref{eq:mc_EF} define the conversion from EV traffic flows to energy flow rate (measured in kW). The formulation captures how individual EV path flows translate into aggregate energy demand and supply rate at each charging node at each FCS. This conversion is essential for coupling the transportation and power networks:
\small{\begin{subequations}\label{eq:mc_EF}
\begin{align}
  &\mathrm{EF}^{+}_{n}
  =\sum_{rs\in\mathcal{S}}\sum_{q\in \mathcal{Q}_{rs}'}      f^{q,\mathrm{EV}}_{rs}\,E^{q,+}_{n'}\,\gamma^{q,+}_{n'},\quad
      \forall \;n\in\mathcal{N}^{\rm fcs}
      \label{eq:mc_EFplus}\\
  &\mathrm{EF}^{-}_{n}
  =\sum_{rs\in\mathcal{S}}\sum_{q\in \mathcal{Q}_{rs}'}
      f^{q,\mathrm{EV}}_{rs}\,E^{q,-}_{n''}\,\gamma^{q,-}_{n''} ,\quad
      \forall \;n\in\mathcal{N}^{\rm fcs}
      \label{eq:mc_EFminus}\\
      &\mathrm{EF}^{+}_{n}+\mathrm{EF}^{-}_{n} 
      \;\leq\;  
      \overline P^{\mathrm{evcs}}\,c_{n},\quad
      \forall \;n\in\mathcal{N}^{\rm fcs}\label{leq:mc_truncate_rho}
\end{align}
\end{subequations}}
\subsection{The Distribution System Operator (DSO)} \label{sec:DSO}
\noindent The DSO manages a radial distribution network characterized by buses $i\in \Omega^B $ and lines $ (i,j) \in \Omega^L$. Acting as a price-taker subject to given nodal wheeling fees and fixed loads, 
the DSO determines the network operating state, including voltage, line flows, and power balances, so as to maximize revenue and ensure AV network feasibility. We formulate the problem as the Linearized DistFlow model, originally proposed in~\cite{baran2002network}, which neglects higher-order loss terms.
\small{
\begin{subequations}\label{eq:DSO}
\begin{align}
&\max_{\Xi_{\rm dso}}
\;
\sum_{i\in\Omega^{B}\setminus{\{i_{\rm tso}\}}} 
w_{i}^{\rm dso} \, \widehat{P}_{i}^{\rm dso}
 \;+\;
(m_i^{\rm tso} - m^{\rm ws})\widehat{P}_{i}^{\rm tso}
\label{eq:dso_obj}
\\
&\text{s.t. } \;U^{\mathrm{sqr}}_{i} - U^{\mathrm{sqr}}_{j} 
    \;{=}\;  
2(r_{ij}p^{\mathrm{line}}_{ij}+x_{ij}q^{\mathrm{line}}_{ij}), 
\quad (i,j)\in\Omega^{L}
       \label{eq:dso_vdrop_alt}
\\
& -S^{\max}_{ij} 
\;{\le}\; p^{\mathrm{line}}_{ij} 
\;{\le}\; S^{\max}_{ij}, \quad
(i,j)\in\Omega^{L}
\label{eq:dso_Plim_alt}
\\
 &-S^{\max}_{ij} 
\;{\le}\; q^{\mathrm{line}}_{ij} 
\;{\le} \; S^{\max}_{ij}, \quad
(i,j)\in\Omega^{L}
    \label{eq:dso_Qlim_alt}
\\
&\underline{U}^{2}
\;{\le} \;U^{\mathrm{sqr}}_{i}
\;{\le}\;
\overline{U}^{2},\quad
i\in\Omega^{B}
    \label{eq:dso_ulim_alt}
\\
&\widehat{P}_{i}^{\rm dso}
    \;{=}\; \sum_{ik \in \Omega^L_{i}} p^{\mathrm{line}}_{ik} - \sum_{ji \in \Omega^L} p^{\mathrm{line}}_{ji}, \;\;
    i \in\Omega^{B}_{-tso}
    \label{eq:dso_Pbal_alt}
\\
&\sum_{i\in\Omega^{B}_{-tso}} \widehat{P}_{i}^{\rm tso}
\;{=}\;
    \sum_{i_{\rm tso}k \in \Omega^L_{i_{\rm tso}}} p^{\mathrm{line}}_{i_{\rm tso}k},
    \label{eq:dso_Pbal_root_alt}
\\
&\underline{P}_{\rm tso}
\;
{\le}\;
\displaystyle\sum_{i\in \Omega^B \setminus 
\{i_{\rm tso}\}}\widehat{P}_{i}^{\rm tso}
\;{\le}\;
\overline{P}_{\rm tso}
\label{eq:tso_active_lim_alt}
\\
&\underline{q}_i 
\;{\leq}\;
q^{\rm load}_{i}
+
\sum_{ik \in \Omega^L_{i}} q^{\mathrm{line}}_{ik }
 - \sum_{ji \in \Omega^L_{i}}q^{\mathrm{line}}_{ji} 
\;{\leq}\;
\overline{q}_{i}, \,
i\in\Omega^{B}_{-tso}
\label{eq:dso_Qbal_alt}
\\
\label{eq:tso_reactive_lim_alt}
&\underline{Q}_{\rm tso}
\;
{\le}\;
\sum_{i_{\rm tso}k \in \Omega^L_{j}} q^{\mathrm{line}}_{i_{\rm tso}k}
\;
{\le}\;
\overline{Q}_{\rm tso}
\end{align}
\end{subequations}}
where $\Xi_{\rm DSO} = 
\{ 
\widehat{P}_i^{\rm dso}, \widehat{P}_i^{\rm tso},  i \in \Omega^{B}_{-i_{\rm tso}};\;
U^{\mathrm{sqr}}_{i}\ge 0,  i \in\Omega^B;
  p^{\mathrm{line}}_{ij}, q^{\mathrm{line}}_{ij},
  I_{ij}\ge 0, (i,j)\in\Omega^{L}
  \}$.
  
\subsection{Competing Load Serving Entities (LSEs)} \label{sec:LSEs}
\noindent 
Each LSE \(f \in \mathcal{F}\) participates in the electricity market by producing and supplying energy through the distribution network. Each LSE operates multiple generation units connected to buses \(i \in \Omega^{B}\) and is modeled as a price-taking entity that anticipate the nodal wheeling fee \(\{w_{i}^{\rm lse}\}_{i \in \Omega^{B}}\), other LSEs' household sales at each demand bus \(i \in \Omega^{B}\), as well as the trading prices with the CNO, \(\{\phi_{f,i}^{\rm cno,\pm}\}\), and that with the DSO, \(\{\phi_{f,i}^{\rm tso}\}\). Each LSE determines its generation output \(\{p^{\mathrm{gen}}_{f,i}\}\) and allocates power sales \(\{p^{\mathrm{sell}}_{f,i}\}\) across demand buses to maximize profit. The LSEs' decisions are subject to generation capacity limits and a LSE-level energy balance between total generation and total sales, as formulated below:

{\begin{subequations}\label{eq:gen}
\begin{align}
&\max_{\Xi_{f} }
\; 
\displaystyle\sum_{i\in\Omega^{B}_{-tso}}
\Bigg[\underbrace{R_{i}\!\bigl(P^{\mathrm{sell}}_{i}\bigr)
\,\widehat{p}^{\mathrm{sell}}_{f,i}}_{\text{households revenue}}
 -
    \underbrace{m_i^{\rm lse} \phi_{f,i}^{\rm tso}}_{\text{TSO supply cost}}
    + \underbrace{M_{i}^{\rm lse,+} \phi_{f,i}^{\rm cno,+}}_{\text{CNO revenue}}
\notag\\
      & 
    -
    \underbrace{M_{i}^{\rm lse,-} \phi_{f,i}^{\rm cno,-}}_{\text{CNO cost}} 
    -
\underbrace{
            w_{i}^{\rm lse}\,\left\{\widehat{p}^{\mathrm{sell}}_{f,i}
            -\widehat{p}^{\mathrm{gen}}_{f,i} 
           + \phi_{f,i}^{\rm cno,+}
     -  \phi_{f,i}^{\rm cno,-}
     \right\}}_{\text{wheeling fee}}
    \Bigg]
      \label{eq:gen_obj}\\
    & \;-\; \displaystyle\sum_{i\in\Omega^G_f}
\underbrace{ C_{f,i}\!\bigl(\widehat{p}^{\mathrm{gen}}_{f,i}\bigr)\widehat{p}_{f,i}^{\rm gen}}_{\text{generation cost}} \;-\; 
\underbrace{\rho \cdot LS}_{\text{load shedding}}
  \notag\\
  &\text{s.t.}\quad
\underline p^{\mathrm{gen}}_{f,i}
\;{\le}\; \widehat{p}^{\mathrm{gen}}_{f,i}
\;{\le}\;
     \overline p^{\mathrm{gen}}_{f,i},
     \qquad\forall i\in\Omega^G_f
     \label{eq:gen_cap}\\
  &  
    \sum_{i\in\Omega^{B}_{-i_{\rm tso}}} \widehat{p}^{\mathrm{sell}}_{f,i}
   - 
   \delta_{f,i}^{\rm gen}
   \widehat{p}^{\mathrm{gen}}_{f,i}
  -\phi_{f,i}^{\rm cno,-}
  +\phi_{f,i}^{\rm cno,+}
  -\phi_{f,i}^{\rm tso}
   \;{=}\;
  0
  \label{eq:gen_bal}\\
  &
   \qquad \underline{p}_i^{\rm demand}\;\leq\;\displaystyle\sum_{f\in\mathcal{F}} \widehat{p}_{f,i}^{\rm sell} + LS
  \label{eq:gen_lb}
\end{align}
\end{subequations}}

\noindent where 
\(\Xi_{f} = \{
    \widehat{p}_{f,i}^{\text{sell}}\geq 0, \;
    \widehat{p}_{f,i}^{\text{gen}}\geq 0, \;
    \phi_{f,i}^{\rm cno, \pm} \geq 0, \;
    \phi_{f,i}^{TSO}\geq 0, \;
    LS \geq 0
\}\), \(P_i^{\rm sell} =  \displaystyle\sum_{f\in\mathcal{F}}\widehat{p}^{\rm sell}_{f,i} \), and $\delta_{f,i}^{\rm gen}$ indicates whether LSE $f$ owns generators at bus $i$. Any differentiable sales quantity and cost functions are sufficient for analysis, For computational purpose, we take \(R_{i}\!\bigl(P^{\mathrm{sell}}_{i}\bigr) =a_{i}P_{i}^{\mathrm{sell}} + b_{i}\) and \(C_{f,i}\!\bigl(\widehat{p}^{\mathrm{gen}}_{f,i}\bigr) = d_{f,i}\widehat{p}^{\mathrm{gen}}_{f,i} + e_{f,i}\).

\subsection{Market Clearing Conditions}\label{sec:mc}
\noindent 
The market-clearing conditions are the equations that consist of shared variables and must be satisfied by all players. These coupling constraints link the decisions of different players (CNO, DSO, LSEs, traffic flows) and ensure that the shared variables have consistent values across all players' optimization problems.

\textbf{DSO Injection Power Balance}
\small{\begin{equation}\label{eq:mc_DSO_power_balance}
\begin{aligned}
\widehat{P}_{i}^{\rm dso}
&
\overset{w_i}{=}
\sum_{f\in\mathcal{F}} (\widehat{p}^{\mathrm{sell}}_{f,i}
-
\widehat{p}^{\mathrm{gen}}_{f,i}
-
\phi_{f,i}^{\rm cno,-}
+
\phi_{f,i}^{\rm cno,+}),  \forall i\in\Omega^{B}_{-tso}
\end{aligned}
\end{equation}}

\textbf{Power Quantity Consistency}
{\small
\begin{subequations}
\begin{align}
&\text{CNO-Traffic:}  
\quad p_{{\rm cno},n}^{\rm fcs,\pm}
\overset{\alpha_n^{\pm}}{=}
EF_{n}^{\pm},
\quad \forall n\in\mathcal{N}^{\rm fcs}
\label{eq:mc_cno_ev}
\\
&\text{CNO-LSE:}
\quad \sum_{n\in\Omega^{\rm fcs}_{i}}
p^{\rm fcs,\pm}_{cno,n}
\overset{M_i^{\pm}}{=}
\sum_{f\in\mathcal{F}}\phi_{f,i}^{\rm cno,\pm},
\quad \forall i\in\Omega^{B}_{-tso}
\label{eq:mc_CNO_LSE_power_balance}
\\
&\text{LSE-DSO:}
\quad \sum_{f\in\mathcal{F}} \phi_{f,i}^{\rm tso}
\overset{m_i^{\pm}}{=}
\widehat{P}_{i}^{\rm tso},
\quad \forall i\in\Omega^{B}_{-tso}
\label{eq:mc_TSO_LSE}
\end{align}
\end{subequations}
}

\textbf{Price Consistency}
\small{
\begin{subequations}\label{eq:mc_price}
\begin{align}
w_{i}^{\rm lse}
&=
w_{i}^{\rm dso}
=
w_i,
\quad \forall i\in\Omega^{B}
\label{eq:mc_w}
\\
\alpha^{\pm}_{{\rm UE},n}
&=
\alpha^{\pm}_{{\rm cno},n}
=
\alpha_n^{\pm},
\quad \forall n\in\Omega_i^{\rm fcs},\; i\in\Omega^{B}
\label{eq:mc_alpha_pms}
\\
M_i^{\rm cno,\pm}
&=
M_i^{\rm lse,\pm}
= M_i^{\pm},
\quad \forall n\in\Omega_i^{\rm fcs},\; i\in\Omega^{B}
\label{eq:mc_M}
\\
m_i^{\rm tso}
&=
m_i^{\rm lse}
= m_i,
\quad \forall  i\in\Omega^{B}_{-\rm tso}
\label{eq:mc_m}
\end{align}
\end{subequations}}

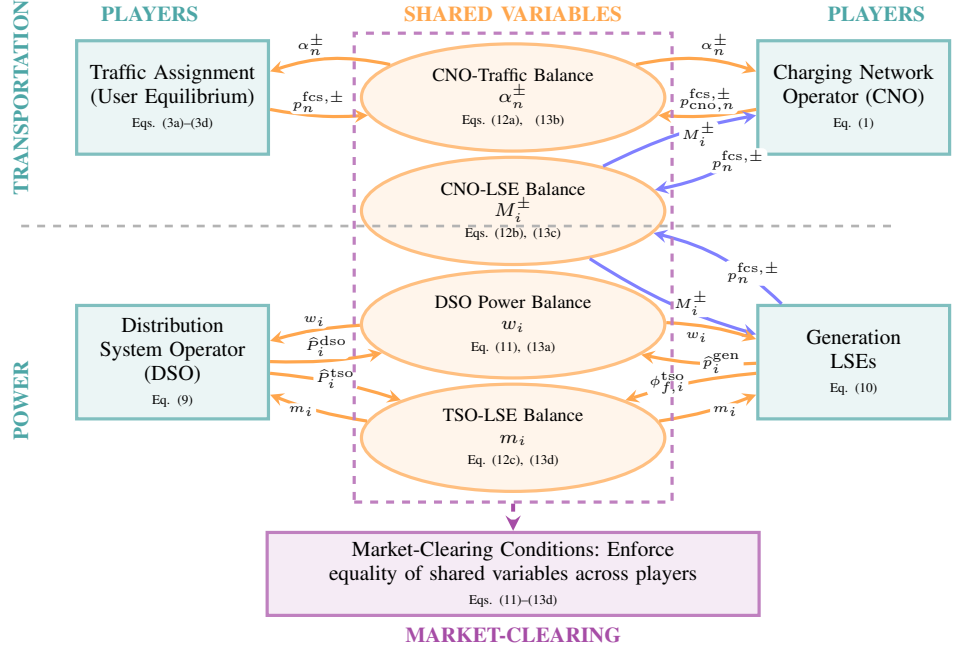
\begin{figure*}[t]
\centering
\begin{tikzpicture}[
    node distance=3cm,
    player/.style={rectangle, draw=teal!50, fill=teal!8, very thick, minimum height=1.5cm, minimum width=2.5cm, text width=2.3cm, align=center, font=\footnotesize},
    shared/.style={ellipse, draw=orange!50, fill=orange!8, very thick, minimum height=1.2cm, minimum width=2.8cm, text width=2.6cm, align=center, font=\scriptsize},
    input/.style={rectangle, draw=olive!60, fill=olive!10, thick, minimum height=0.6cm, minimum width=1.5cm, text width=1.3cm, align=center, font=\scriptsize},
    arrow/.style={thick, -stealth, line width=1.2pt},
    label/.style={font=\footnotesize\bfseries, text=black},
    varlabel/.style={font=\tiny, text=black, inner sep=1pt, fill=white, rounded corners=2pt}
]

\node[label] at (-1.8,6.6) {\textcolor{teal!70}{PLAYERS}};
\node[label] at (7.8,6.6) {\textcolor{teal!70}{PLAYERS}};
\node[label] at (3,6.6) {\textcolor{orange!70}{SHARED VARIABLES}};
\node[label] at (3,-1.6) {\textcolor{violet!70}{MARKET-CLEARING}};

\node[player] (traffic) at (-1.5,5.5) {Traffic Assignment\\(User Equilibrium)\\{\tiny Eqs.~\eqref{eq:total_flow}--\eqref{fv:user_equilibrium}}};
\node[player] (cno) at (7.5,5.5) {Charging Network\\Operator (CNO)\\{\tiny Eq.~\eqref{eq:CNO_basic}}};
\node[player] (dso) at (-1.5,2) {Distribution\\System Operator\\(DSO)\\{\tiny Eq.~\eqref{eq:DSO}}};
\node[player] (gen) at (7.5,2) {Generation\\LSEs\\{\tiny Eq.~\eqref{eq:gen}}};

\node[shared] (shared0) at (3,5.5) {CNO-Traffic Balance\\$\alpha_n^{\pm}$\\{\tiny Eqs.~\eqref{eq:mc_cno_ev},
~\eqref{eq:mc_alpha_pms}}};
\node[shared] (shared1) at (3,4) {CNO-LSE Balance\\$M_i^{\pm}$\\{\tiny Eqs.~\eqref{eq:mc_CNO_LSE_power_balance},~\eqref{eq:mc_M}}}; 
\node[shared] (shared2) at (3,2.5) {DSO Power Balance\\$w_i$\\{\tiny Eq.~\eqref{eq:mc_DSO_power_balance},~\eqref{eq:mc_w}}}; 
\node[shared] (shared3) at (3,1) {TSO-LSE Balance
\\$m_i$\\{\tiny Eq.~\eqref{eq:mc_TSO_LSE},~\eqref{eq:mc_m}}
}; 

\node[rectangle, draw=violet!50, fill=violet!8, very thick, minimum height=1cm, minimum width=6.5cm, text width=5cm, align=center, font=\footnotesize] (mc) at (3,-0.8) {Market-Clearing Conditions: Enforce equality of shared variables across players\\{\tiny Eqs.~\eqref{eq:mc_DSO_power_balance}--\eqref{eq:mc_m}}}; 


\draw[arrow, orange!70] (shared0) to[bend right=15] node[varlabel, above left] {$\alpha_{n}^{\pm}$} (traffic);
\draw[arrow, orange!70] (traffic) to[bend right=7] node[varlabel, above] {$p_{n}^{\mathrm{fcs},\pm}$} (shared0);

\draw[arrow, orange!70] (shared0) to[bend left=15] node[varlabel, above right] {$\alpha_{n}^{\pm}$} (cno);
\draw[arrow, orange!70] (cno) to[bend left=7] node[varlabel, above] {$p_{{\rm cno},n}^{\mathrm{fcs},\pm}$} (shared0);

\draw[arrow, blue!50] (shared1) to[bend left=8] node[varlabel, right] {$M_i^{\pm}$} (cno);
\draw[arrow, blue!50] (cno) to[bend left=10] node[varlabel, above right] {$p_{n}^{\mathrm{fcs},\pm}$} (shared1);

\draw[arrow, blue!50] (shared1) to[bend right=8] node[varlabel, right] {$M_i^{\pm}$} (gen);
\draw[arrow, blue!50] (gen) to[bend right=15] node[varlabel, below right] {$p_n^{\rm fcs,\pm}$} (shared1); 

\draw[arrow, orange!70] (shared2) to[bend right=6] node[varlabel, above] {$w_i$} (dso);
\draw[arrow, orange!70] (dso) to[bend right=6] node[varlabel, above] {$\widehat{P}_{i}^{\rm dso}$} (shared2);

\draw[arrow, orange!70] (shared2) to[bend left=7] node[varlabel, below left] {$w_i$} (gen);
\draw[arrow, orange!70] (gen) to[bend left=7] node[varlabel, right] {$\widehat{p}_{i}^{\mathrm{gen}}$} (shared2);

\draw[arrow, orange!70] (shared3) to[bend left=6] node[varlabel, left] {$m_i$} (dso);
\draw[arrow, orange!70] (dso) to[bend left=6] node[varlabel, right, pos=0.3] {$\widehat{P}_{i}^{\rm tso}$} (shared3);

\draw[arrow, orange!70] (shared3) to[bend right=6] node[varlabel, right] {$m_i$} (gen);
\draw[arrow, orange!70] (gen) to[bend right=6] node[varlabel, left] {$\phi_{f,i}^{\rm tso}$} (shared3); 

\draw[dashed, gray!60, line width=1pt] (-3.5,3.8) -- (8.,3.8);
\node[label, rotate=90] at (-3.5,5.5) {\textcolor{teal!70}{TRANSPORTATION}};
\node[label, rotate=90] at (-3.5,1.5) {\textcolor{teal!70}{POWER}};

\node[rectangle, draw=violet!50, dashed, very thick, minimum height=6.2cm, minimum width=4.2cm, fill=none] (shared_box) at (3,3.25) {};

\draw[arrow, violet!70, dashed, line width=1.5pt] (shared_box.south) -- (mc.north);

\end{tikzpicture}
\caption{Player interactions and shared variables in the integrated power-transportation system. Orange ellipses represent shared variables enforced to be equal through market-clearing conditions (violet box). Arrows show dependency directions: players contribute to shared variables they influence, and receive shared variables as inputs to their optimization problems. Blue arrows highlight the inter-network coupling.}
\label{fig:player_interactions}
\end{figure*}

\vspace{-0.1in}

\section{Equilibrium and VI Formulations}\label{sec:overall problem}

\noindent We define an equilibrium solution as a primal-dual point at which the optimality (Karush-Kuhn-Tucker) conditions of all submodules, namely the CNO, transportation network, DSO, and LSEs, together with the market-clearing conditions, are simultaneously satisfied.
In the equilibrium analysis, the price consistency conditions in \eqref{eq:mc_price} are imposed, ensuring that all market participants adopt identical prices for the same transactions. We use the common notation $\{w_i\}$, $\{\alpha_n^{\pm}\}$, $\{M_i^{\pm}\}$, and $\{m_i^{\pm}\}$ throughout the remainder of the paper.
All simplified variables are summarized in compact form in Table~\ref{tab:variables}.

\begin{table}[!t]
\caption{Definition of Variables in VI}
\label{tab:variables}
\footnotesize
\centering
\setlength{\tabcolsep}{3pt}
\renewcommand{\arraystretch}{0.95}
\begin{tabularx}{\columnwidth}{@{}l X@{}}
\toprule
\textbf{Symbol} & \textbf{Definition} \\
\midrule

$\mathbf{v}$ 
& $\triangleq [\mathbf{v}_{\rm cno};\, \mathbf{v}_{\rm traffic};\, \mathbf{v}_{\rm dso};\, \mathbf{v}_{\rm lse}]$ \\

$\mathbf{v}_{\rm cno}$ 
& $\triangleq [p_n^{\rm fcs,\pm}]_{i \in \Omega^{B}_{-tso},\, n \in \Omega_i^{\rm fcs}}$ \\

$\mathbf{v}_{\rm traffic}$ 
& $\triangleq [f_{rs}^{q',\rm EV},\, f_{rs}^{q,\rm FV}]_{rs\in\mathcal{S},\, q'\in \mathcal{Q}'_{rs},\, q\in \mathcal{Q}_{rs}}$ \\

$\mathbf{v}_{\rm dso}$ 
& $\triangleq [\widehat{P}_{i}^{\rm dso},\, \widehat{P}_{i}^{\rm tso},\, U_j^{\rm sqr},\, p_{i'j'}^{\rm line},\, q_{i'j'}^{\rm line}]_{i\in\Omega^{B}_{-tso},\, j\in\Omega^{B}_{-tso},\, (i',j')\in\Omega^L}$ \\

$\mathbf{v}_{\rm lse}$ 
& $\triangleq [\widehat{p}_{f,i}^{\rm sell},\, \widehat{p}_{f,i}^{\rm gen},\, \phi_{f,i}^{\rm cno,\pm},\, \phi_{f,i}^{\rm tso}]_{f\in\mathcal{F},\, i\in\Omega^{B}_{-tso}}$ \\

\midrule

$\mathbf{y}$ 
& $\triangleq [\mathbf{w};\, \mathbf{M};\, \mathbf{m};\, \boldsymbol{\alpha}]$ \\

$\mathbf{w}$ 
& $\triangleq [w_i]_{i \in \Omega^{B}_{-tso}}$ 
\hspace{0.5in}
$\mathbf{M}$ 
 $\triangleq [M_i^{\pm}]_{i \in \Omega^{B}_{-tso}}$ \\
$\mathbf{m}$ 
& $\triangleq [m_i]_{i \in \Omega^{B}_{-tso}}$ 
\hspace{0.5in}
$\boldsymbol{\alpha}$ 
 $\triangleq [\alpha_n^{\pm}]_{i \in \Omega^{B}_{-tso},\, n\in\Omega_i^{\rm fcs}}$ \\

\bottomrule
\end{tabularx}
\end{table}

Then the overall equilibrium problem can be written as the variational inequality (VI) of finding a pair of primal tuple 
$\overline{\mathbf{v}},\overline{\mathbf{y}}$ satisfying
\begin{equation}\label{eq:overall_VI}
F(\overline{\mathbf{v}},\overline{\mathbf{y}})^\top
\left(\mathbf{v}-\overline{\mathbf{v}}\right)\ge 0,
\quad \forall\,\mathbf{v}\in\mathbb{V},
\qquad
\Psi(\overline{\mathbf{v}},\overline{\mathbf{y}})=0,
\end{equation}
where $\mathbb{V}$ is the (set) Cartesian product concatenating 
all the constraints in the submodules, 
\begin{equation}\label{eq:F_compact}
F(\mathbf{v},\mathbf{y})
=
\big[
F_{\rm cno};
F_{\rm traffic};
F_{\rm dso};
F_{\rm lse}
\big],
\end{equation}

\noindent
where each component of $F$ represents the partial gradient of the objective function with respect to the primary variables
of the corresponding module, i.e., CNO, traffic, DSO, and LSE subproblems, and
\begin{equation}\label{eq:Psi_compact}
\Psi(\mathbf{v},\mathbf{y})
=
\big[
\Psi_{\mathbf{w}};
\Psi_{\mathbf{M}};
\Psi_{\mathbf{m}};
\Psi_{\boldsymbol{\alpha}}
\big],
\end{equation}
\noindent
where $\Psi$ collects the residuals of the market-clearing conditions. Specifically, each component is obtained by rearranging the corresponding market-clearing equations~\eqref{eq:mc_DSO_power_balance}--\eqref{eq:mc_TSO_LSE} to one side.

\noindent Moreover, the mappings \(F(\mathbf{v},\mathbf{y})\) and \(\Psi(\mathbf{v},\mathbf{y})\) admit the structure
\begin{equation}\label{eq:F_Psi_structure}
F(\mathbf{v},\mathbf{y})=-\mathbf{A}^\top\mathbf{m}+H(\mathbf{v}),\;
\Psi(\mathbf{v},\mathbf{y})=\mathbf{A}\mathbf{v}
\end{equation}
where $H(\mathbf{v})
$ is a continuous function in $\mathbf{v}$
and the coupling matrix \(\mathbf{A}\) is a sparse signed incidence matrix that collects the linear market-clearing and consistency conditions. 
It is easy to see that \eqref{eq:overall_VI} is equivalent to the standard VI $(H, \tilde{\mathbb{V}})$, where 
$\tilde{\mathbb{V}} \triangleq \mathbb{V}\cap\{\mathbf{A}\mathbf{v}=0\}$.
The set \(\tilde{\mathbb{V}}\) is convex and compact, and \(H(\mathbf{v})\) is continuous.   From well-known VI theory~\cite{facchinei2003finite}, it follows that the model has at least one equilibrium solution, provided that \(\tilde{\mathbb{V}}\) is nonempty, i.e., if the constrains are altogether consistent.
We can write $\tilde{\mathbb{V}}$ in the compact form $\{\mathbf{v} \geq 0\,\mid\,\mathbf{B}_1\mathbf{v} = \mathbf{b}_1\,,\, \mathbf{B}_2\mathbf{v} \leq \mathbf{b}_2\}$. Therefore, VI$(H, \tilde{\mathbb{V}})$ is equivalent to the following Mixed Complementarity Problem (MCP):
\begin{equation}\label{eq:mcp_compact}
 b_1 - B_1\mathbf{v}  = 0  ;\; 
 0 \le 
\begin{bmatrix} 
\mathbf{v} \\ 
\boldsymbol{\mu} 
\end{bmatrix} 
\perp 
\begin{bmatrix} 
H(\mathbf{v}) + B_1^\top \boldsymbol{\lambda} + B_2^\top \boldsymbol{\mu} \\ 
b_2 - B_2\mathbf{v} 
\end{bmatrix} 
\ge 0
\end{equation}
where $\perp$ is the complementarity notation.

\vspace{-0.1in}
\section{Numerical Analysis}\label{sec:numerical}

\noindent  
This section presents numerical experiments designed to evaluate the proposed equilibrium modeling framework for coupled power and transportation systems with V2G integration. We first describe the test networks and the three operational scenarios considered in Subsection~\ref{sec:netwrok_case}, then examine power network performance in terms of  distribution locational marginal prices (DLMPs) and load shedding outcomes in Subsection~\ref{sec:DLMP_LS}, transportation side EV routing and charging/discharging behavior in Subsection~\ref{sec:ev_bahevior}, and overall system costs in Subsection~\ref{sec:overall_cost}. 
All models were implemented in Julia and solved using the PATH 
solver \cite{DirskeFerris95} 
for the resulting MCP~\eqref{eq:mcp_compact}.

\vspace{-0.1in}

\subsection{Test Networks and Scenarios Design}
\label{sec:netwrok_case}
\begin{figure}[!t]
\centering

\includegraphics[width=0.9\linewidth]{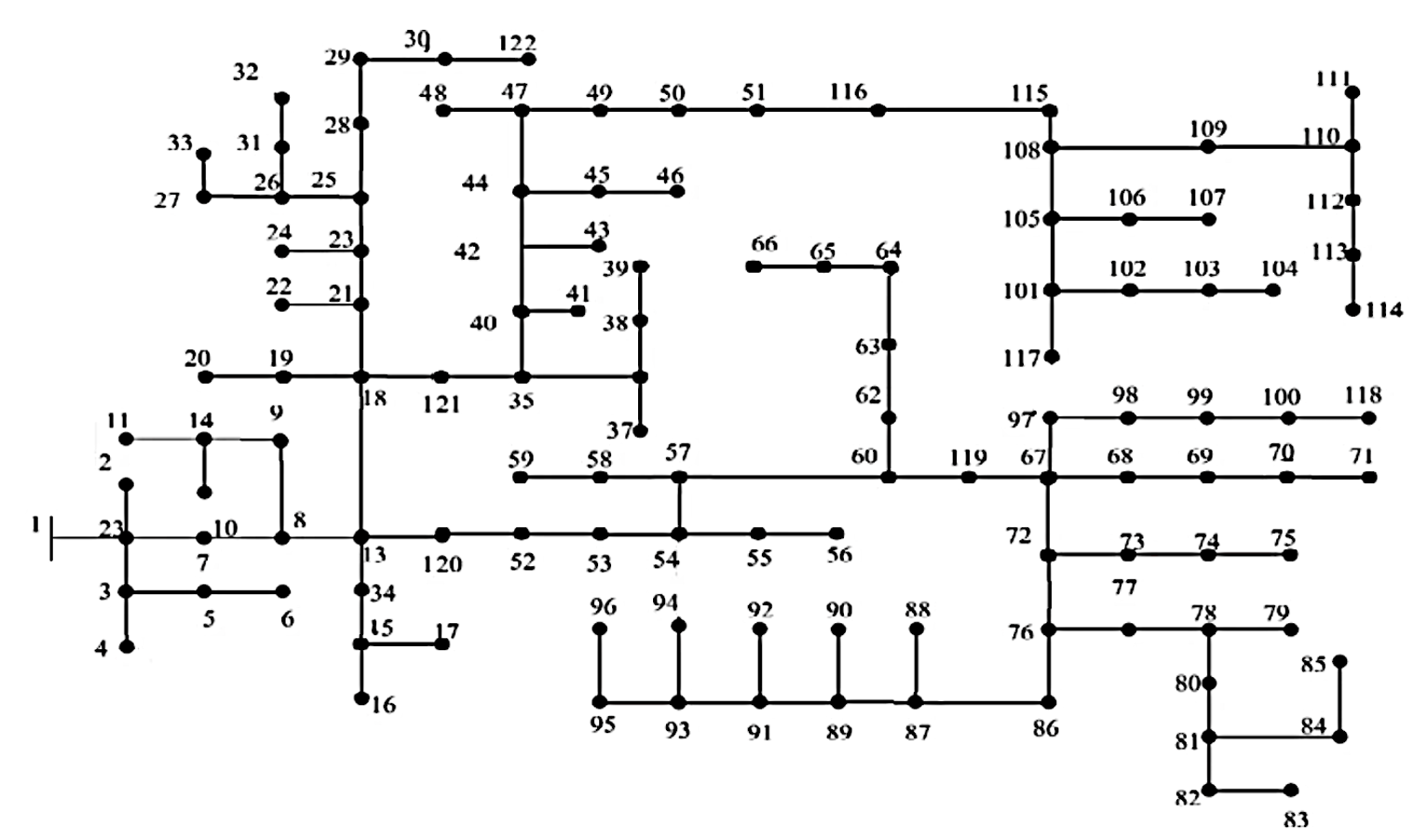}
\caption{Power network: IEEE-123 feeder~\cite{ieee123bus}}
\label{fig:power_network}
\end{figure}

\begin{figure}[!t]
\centering
\includegraphics[width=0.78\linewidth]{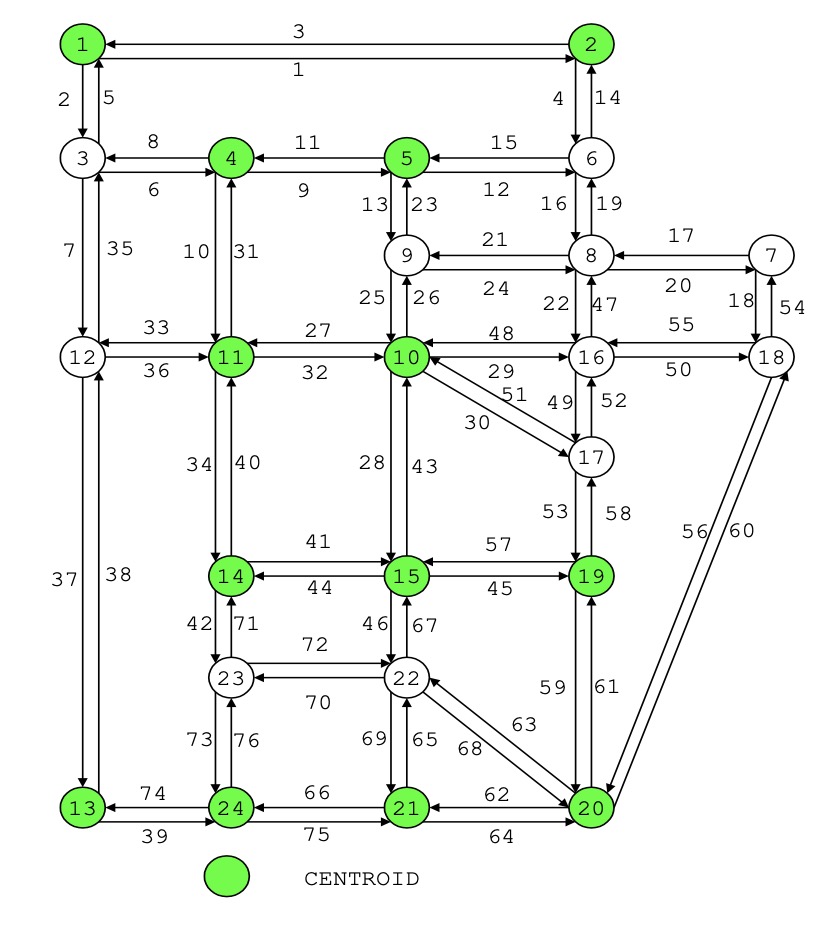}
\caption{Transportation network: Sioux Falls~\cite{bstablerGitHub}}
\label{fig:transportation_network}
\end{figure}

\noindent 
The integrated system comprises a power distribution network and a transportation network. For the power system, we employ the standard IEEE 123-bus test feeder~\cite{ieee123bus}, which represents a radial distribution network with unbalanced loads and voltage regulators, shown in Figure~\ref{fig:power_network}. The feeder supplies a mix of residential loads and FCS demand. Under V2G operation, an FCS may function as a supplier when local discharging amount exceeds charging demand. A TSO connection is modeled at a slack bus, providing upstream power supply at a predetermined wholesale price.
For the transportation network, we adopt the Sioux Falls network~\cite{bstablerGitHub}, consisting of 24 nodes and 76 directed links, shown in Figure~\ref{fig:transportation_network}. Travel demand is fixed and restricted to a subset of OD pairs: origin $\{1, 2, 4, 7, 9\}$ and destinations $\{13, 19, 20, 23, 24\}$. FCSs are located at selected nodes $\{3, 6, 8, 11, 12, 18\}$

\noindent 
To assess the value of V2G under different grid stress conditions, we define three operational scenarios.
\textbf{(i) Base case:} normal operation conditions with typical household load profile. 
\textbf{(ii) Stress case:} a 50\% increase in household load across all buses. This scenario mimics a high demand period. 
\textbf{(iii) Island case:} line outage contingencies disconnect some buses from the TSO, forming three isolated islands. A simultaneous 50\% increase in household load is imposed to all buses. 
For each scenario, we compute two equilibrium solutions: one \textbf{without} V2G capability and one \textbf{with} V2G enabled. 

\vspace{-0.1in}

\subsection{Power System Evaluation}\label{sec:DLMP_LS}
\begin{figure}[!t]
\centering
\begin{subfigure}{1\linewidth}
\centering
\includegraphics[width=0.8\linewidth]{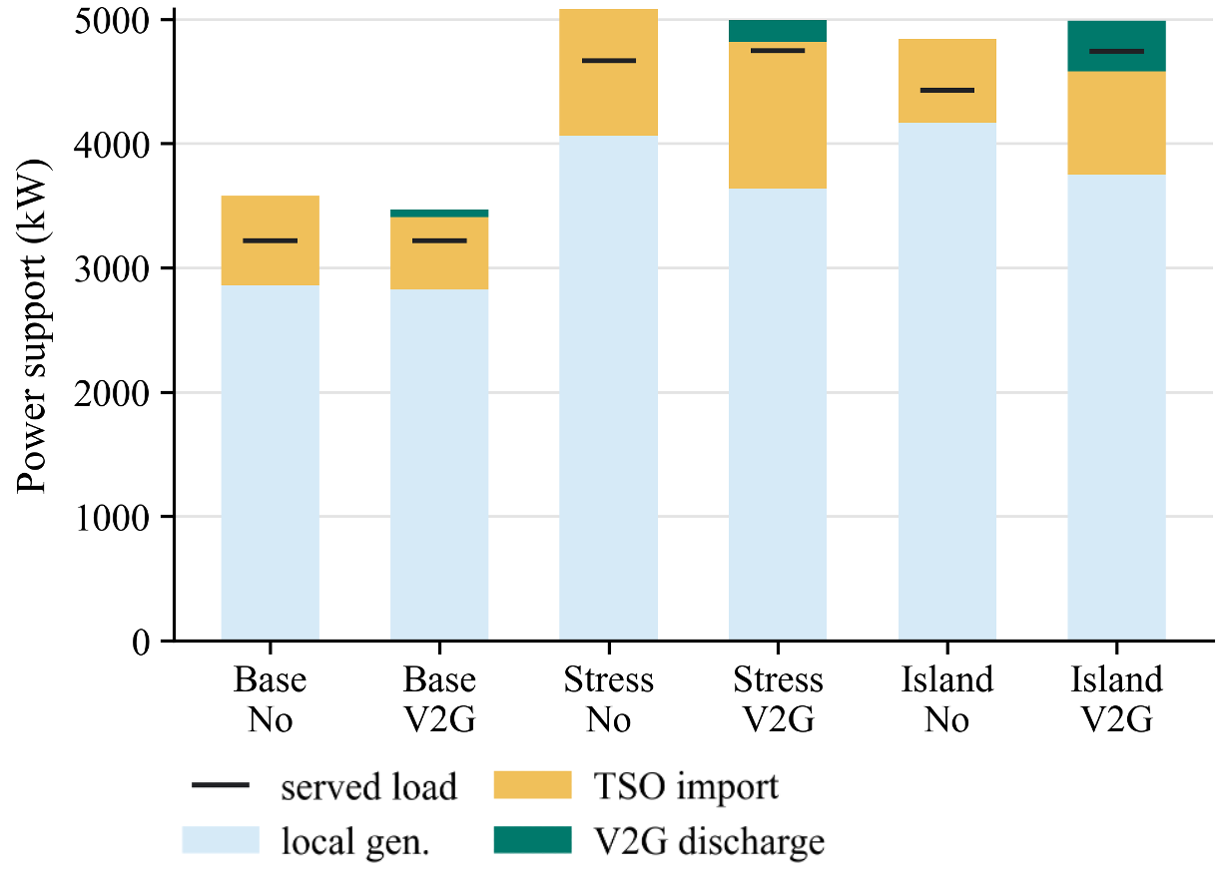}
\caption{Power supply}
\label{fig:supply_and_station_balance_a}
\end{subfigure}

\hfill

\begin{subfigure}{1\linewidth}
\centering
\includegraphics[width=0.8\linewidth]{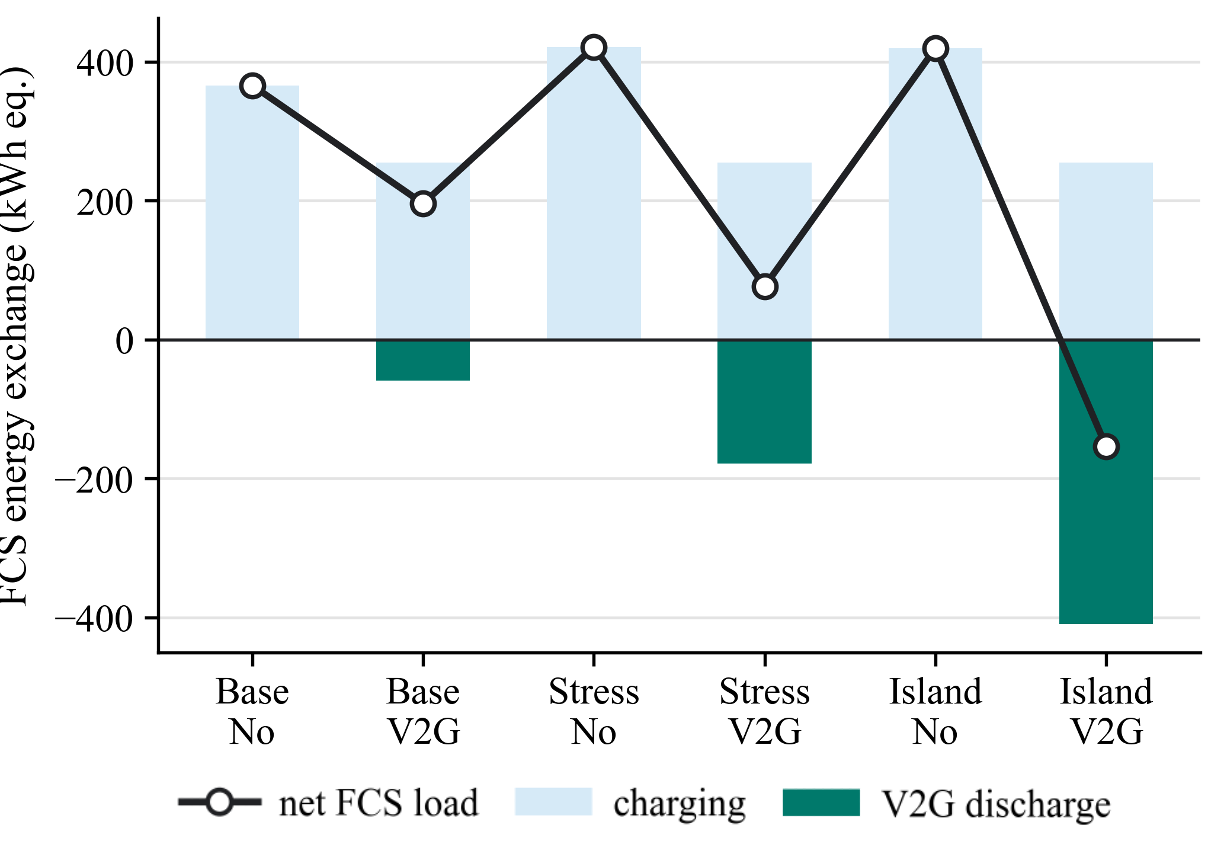}
\caption{Station net load}
\label{fig:supply_and_station_balance_b}
\end{subfigure}

\caption{Power network supply and FCS net load}
\label{fig:supply_and_station_balance}
\end{figure}

\noindent 
We assess power system resilience across varying operational conditions by first examining power supply and station-level energy exchange, followed by their impacts on system scarcity and price signals.
Figure~\ref{fig:supply_and_station_balance_a} illustrates the system's supply-side support. 
In the \textbf{Stress} case and \textbf{Island} case, overall system demand increases, prompting local generation to significantly higher levels. When V2G is activated, EV discharging provides critical supplementary power, transitioning the FCS into a flexible prosumer (Fig.~\ref{fig:supply_and_station_balance_b}) 
This is especially vital in the \textbf{Island} case where accessible local generation and TSO imports are deficient.
Here, the V2G discharge volume exceeds charging demand, driving the station's net load below zero and allowing the FCS to act as an emergency power supplier to support the isolated network.

\begin{figure}[!t]
\centering
\begin{subfigure}{1\linewidth}
\centering
\includegraphics[width=0.93\linewidth]{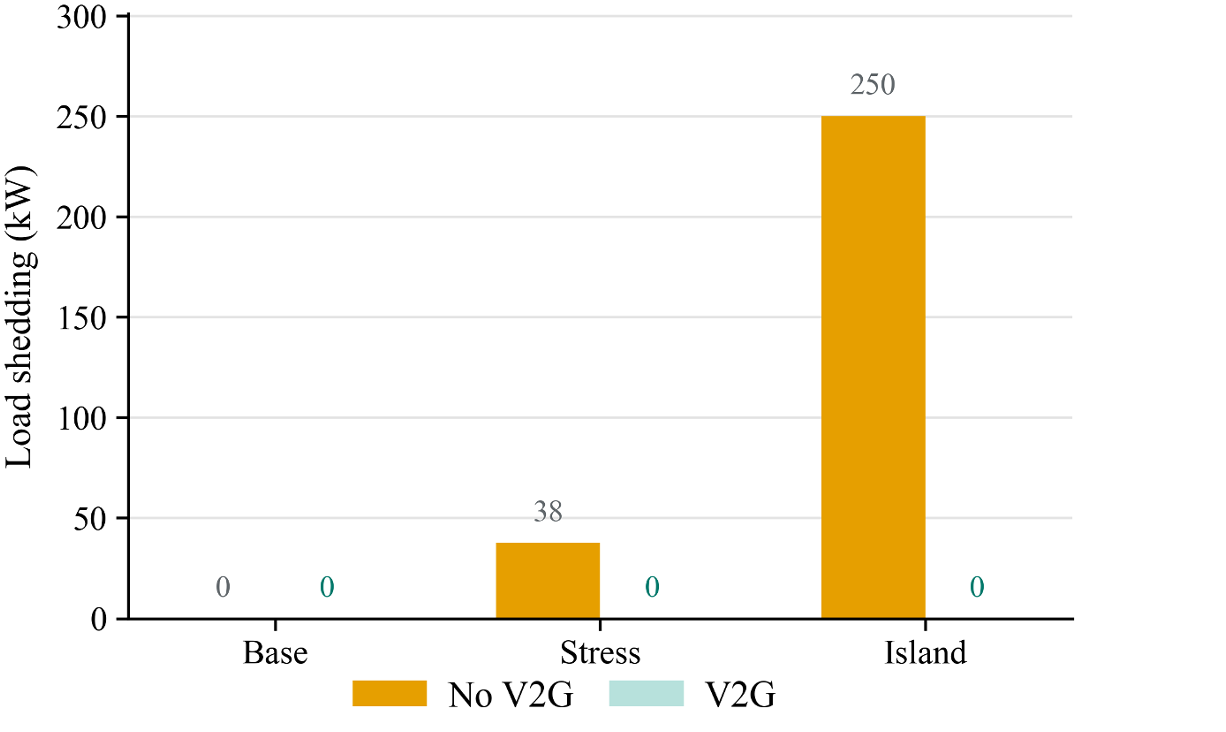}
\caption{Load Shedding (LS)}
\label{fig:scarcity_and_price_a}
\end{subfigure}

\hfill
\begin{subfigure}{1\linewidth}
\centering
\includegraphics[width=0.83\linewidth]{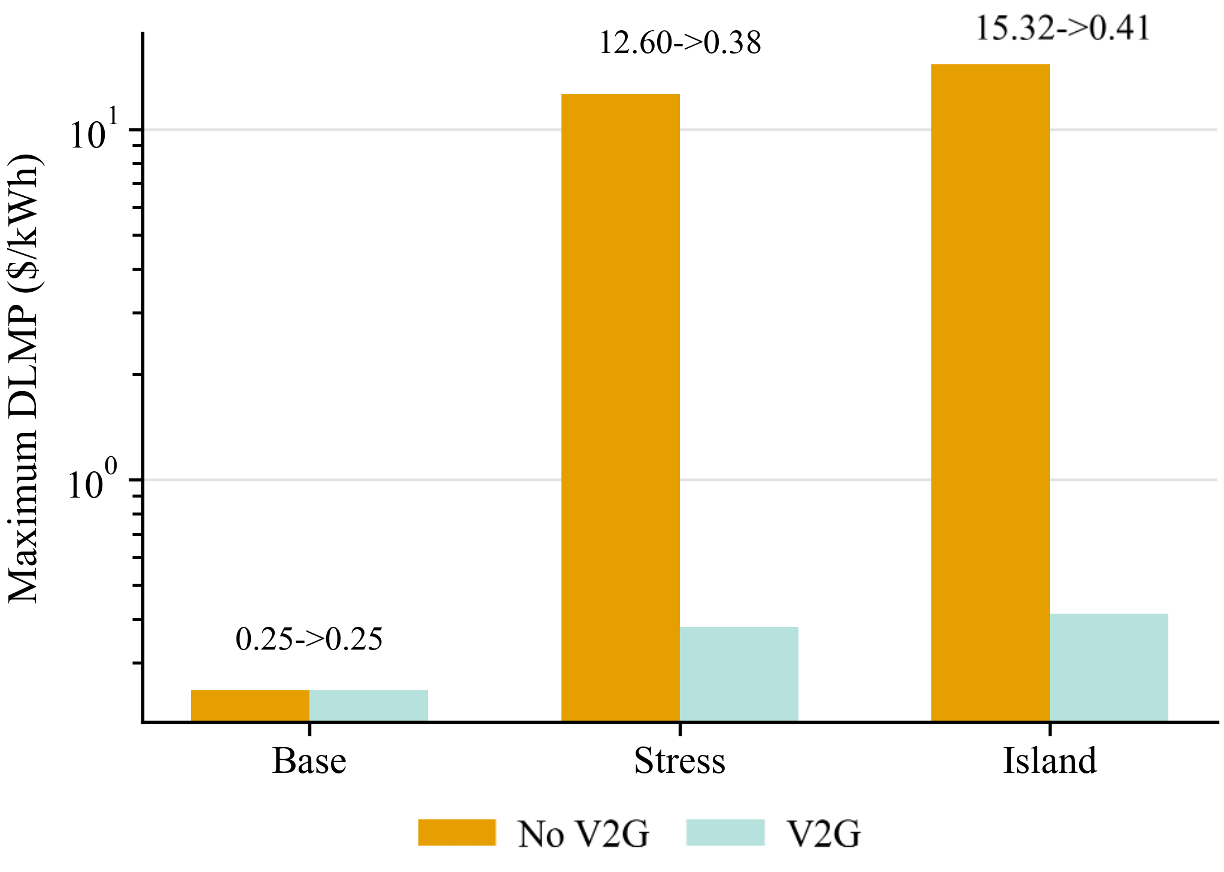}
\caption{Distribution locational marginal price (DLMP)}
\label{fig:scarcity_and_price_b}
\end{subfigure}

\caption{Scarcity quantity and price tail}
\label{fig:scarcity_and_price}
\end{figure}

\noindent To further evaluate system resilience, we introduce two key metrics used to quantify supply adequacy and localized scarcity: Load Shedding (LS) and Distribution Locational Marginal Prices (DLMPs), as shown in Figure~\ref{fig:scarcity_and_price}. 
In the \textbf{Base} case, the system maintains full supply adequacy with zero LS, and the maximum DLMP remains stable at normal levels regardless of V2G integration. Conversely, without V2G, the power system experiences severe scarcity during both the \textbf{Stress} and \textbf{Island} scenarios, indicated by positive LS and exceptionally high peak DLMPs.
Enabling V2G fully mitigates these capacity deficits, reducing LS to zero and driving the maximum DLMPs back to near-normal levels. Consequently, these results demonstrate that V2G-capable EVs serve as a highly effective distributed resource for enhancing overall system resilience, ensuring supply adequacy, and stabilizing prices during critical grid disruptions.

\vspace{-0.1in}

\subsection{EV Behavior}\label{sec:ev_bahevior}
\begin{figure}[!t]
\centering

\begin{subfigure}{0.75\linewidth}
\centering
\includegraphics[width=\linewidth]{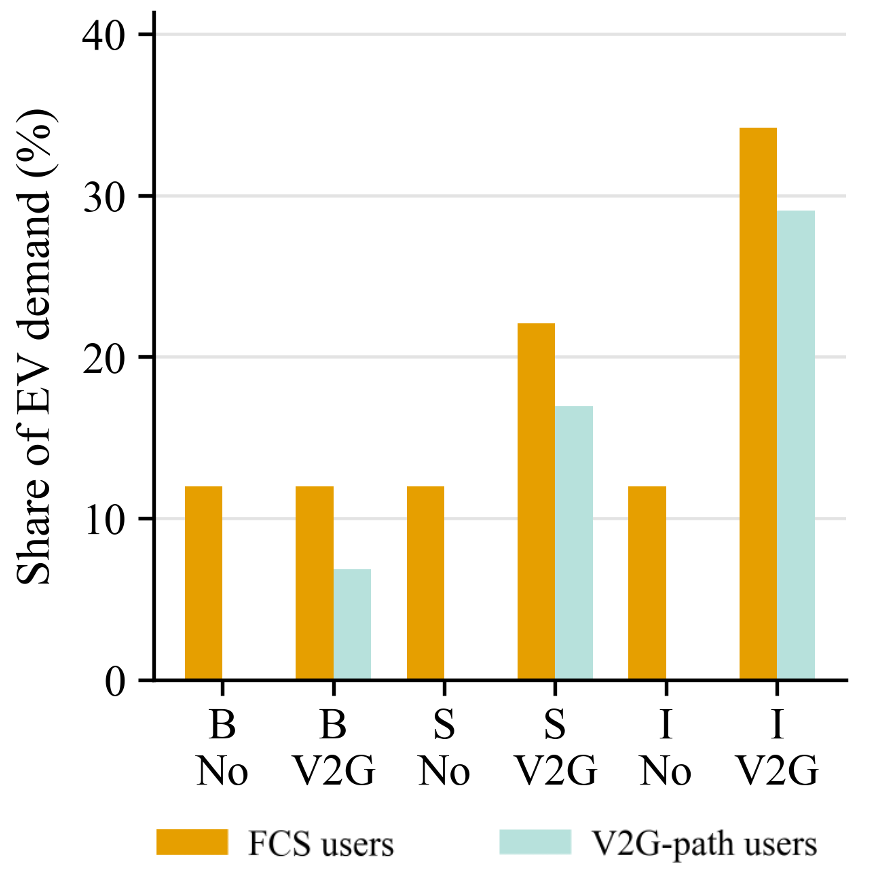}
\caption{Station participation}
\label{fig:ev_fcs_profile_a}
\end{subfigure}
\hfill

\begin{subfigure}{0.77\linewidth}
\centering
\includegraphics[width=\linewidth]{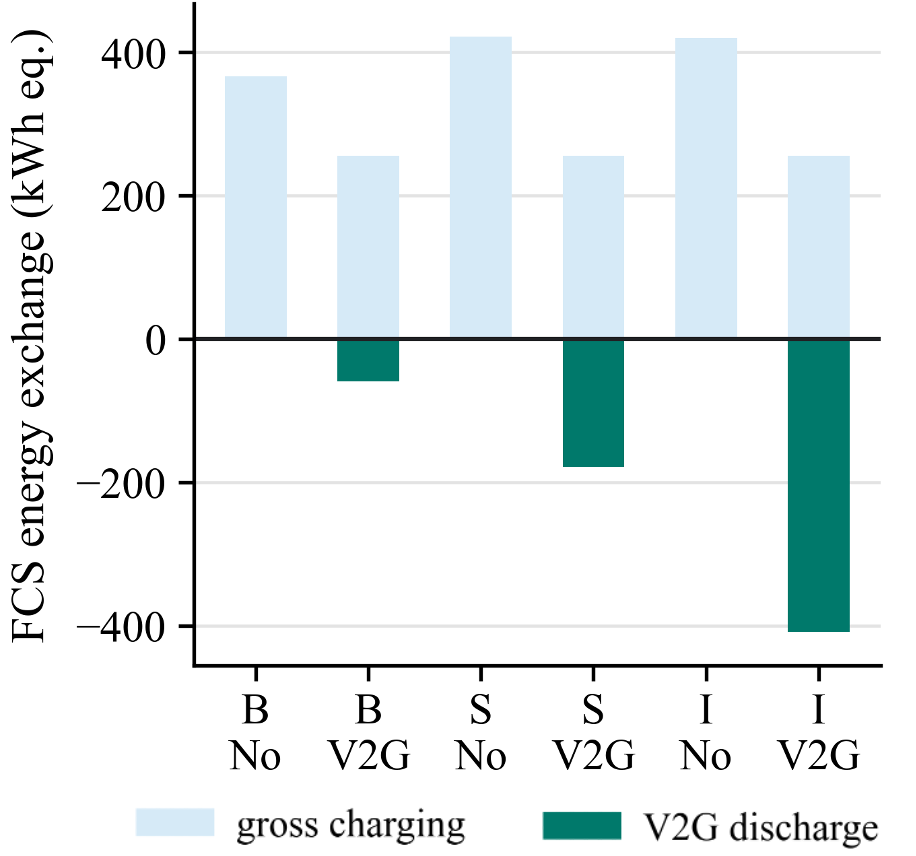}
\caption{Charging and discharging}
\label{fig:ev_fcs_profile_b}
\end{subfigure}
\hfill
\begin{subfigure}{0.75\linewidth}
\centering
\includegraphics[width=\linewidth]{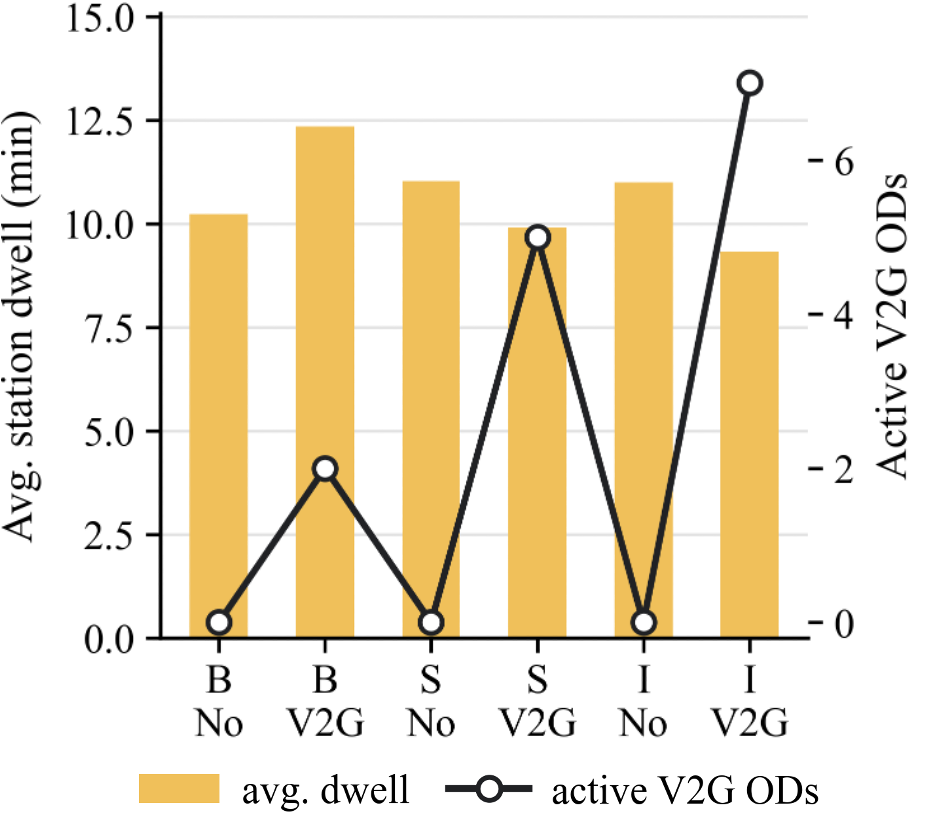}
\caption{Dwell and V2G coverage}
\label{fig:ev_fcs_profile_c}
\end{subfigure}

\caption{EV charging and discharging profile}
\label{fig:ev_fcs_profile}
\end{figure}

\noindent We now examine the transportation-side outcomes, focusing on the charging and discharging behaviors of EVs across the three operational scenarios. Figure~\ref{fig:ev_fcs_profile} summarized key metrics, including the percentages of EV trips utilize FCSs and adopt V2G-capable path, respectively, the gross charging/discharging volume, total waiting and charging/discharging time, and active V2G OD pairs. 
The results indicate a sharp divergence in EV behavior between normal and scarcity conditions. 
Subfigure~\ref{fig:ev_fcs_profile_a}
shows that FCS visitation rates remain unchanged across all No‑V2G cases and in the \textbf{Base} V2G case, but rise to around 22\% and 34\% in stressed and island cases, respectively. In parallel, the usage of discharge-enabled paths is substantially higher in the \textbf{Stress} and \textbf{Island} V2G cases than in the \textbf{Base} V2G case. These sharp increases confirm that scarcity conditions and the loss of upstream supply create strong economic incentives for EVs to engage in discharging.
Subfigure~\ref{fig:ev_fcs_profile_b} 
indicates that V2G reduces gross charging demand in all scenarios, while discharging volume is higher under stress and peaks under island conditions, confirming that significant discharging is triggered under these two abnormal scenarios. 
Subfigure~\ref{fig:ev_fcs_profile_c} reveals that V2G reduces station dwell time in Stress and Island cases but increases in Base case, and the number of discharging OD pairs is larger in stress scenario and greatest in island scenario.

\begin{figure}[!t]
\centering

\begin{subfigure}{1\linewidth}
\centering
\includegraphics[width=0.85\linewidth]{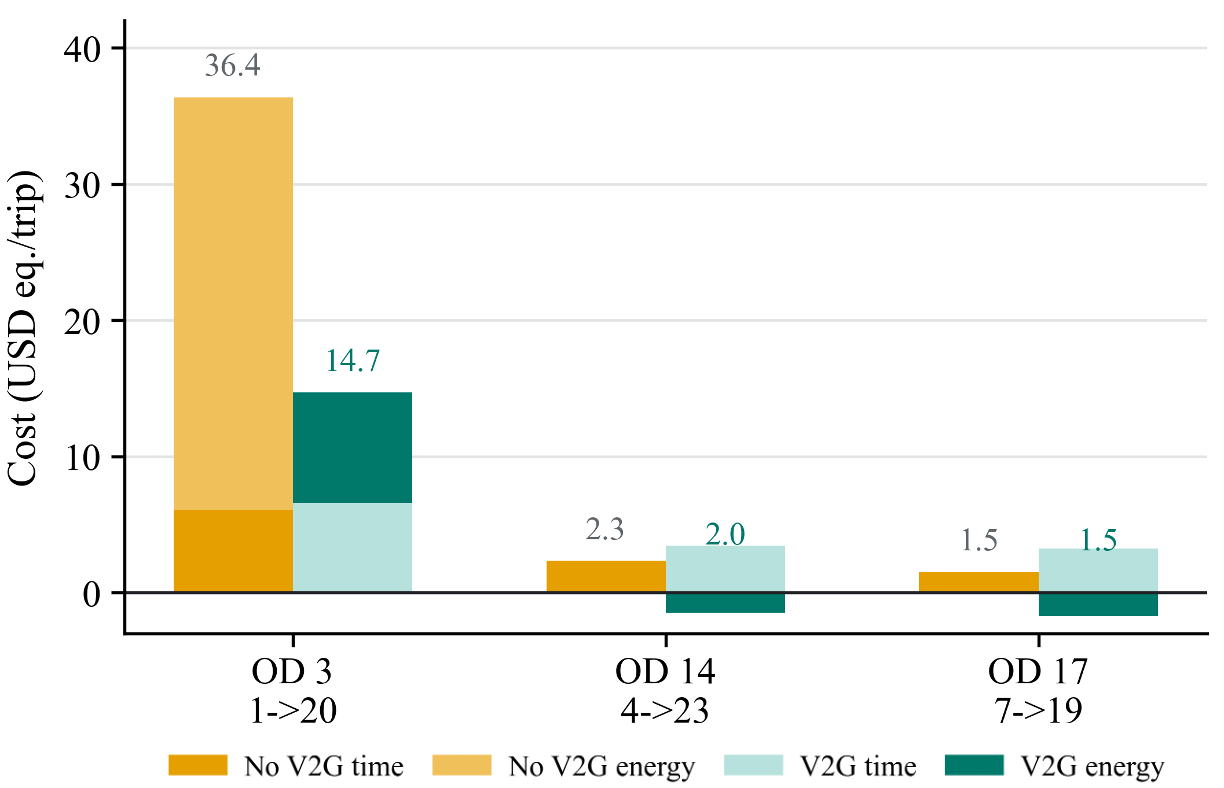}
\caption{Stress dominant-path cost}
\label{fig:dominant-path cost_a}
\end{subfigure}

\hfill
\begin{subfigure}{1\linewidth}
\centering
\includegraphics[width=0.85\linewidth]{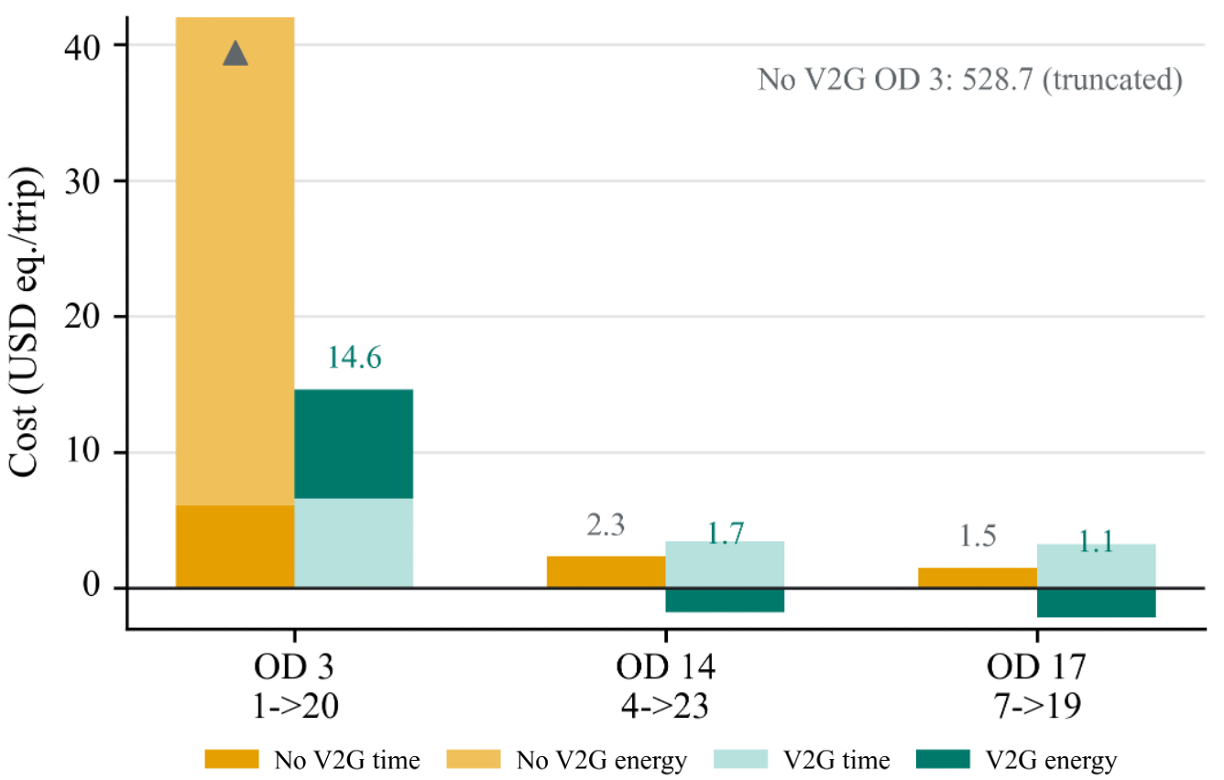}
\caption{Island dominant-path cost}
\label{fig:dominant-path cost_b}
\end{subfigure}

\caption{Generalized path cost}
\label{fig:dominant-path cost}
\end{figure}

\noindent To capture the operational trade-offs in routing decisions, we evaluate generalized path costs for dominant paths used in the \textbf{Stress} and \textbf{Island} cases, shown in Figure~\ref{fig:dominant-path cost}. 
By discharging energy to the grid, EVs generate net revenues (indicated by negative energy costs) that significantly reduce total trip expenses. This is most evident for OD 3, where V2G reduces extreme scarcity-driven costs from \$36.4 (Stress) and \$528.7 (Island) down to approximately \$14.7. Even for less severely impacted routes like OD 14 and OD 17, V2G energy revenues effectively offset time costs, yielding net savings (nonnegative) in all evaluated scenarios.
Ultimately, these reduced path costs provide a compelling economic incentive for EVs to voluntarily reroute their trips and actively participate in grid-supporting discharge events.

\begin{figure}[!t]
\centering

\begin{subfigure}{1\linewidth}
\centering
\includegraphics[width=0.66\linewidth]{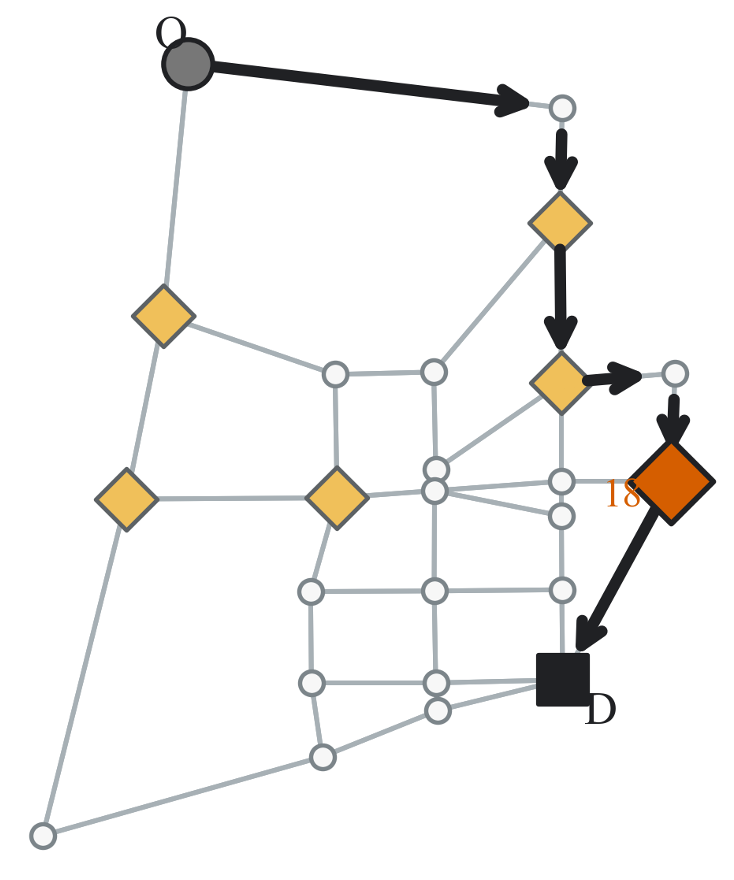}
\caption{No V2G: 22km, 36.4 USD eq. Path:1-2-6-8-7-18-20}
\label{fig:reroute_a}
\end{subfigure}

\hfill
\begin{subfigure}{1\linewidth}
\centering
\includegraphics[width=0.76\linewidth]{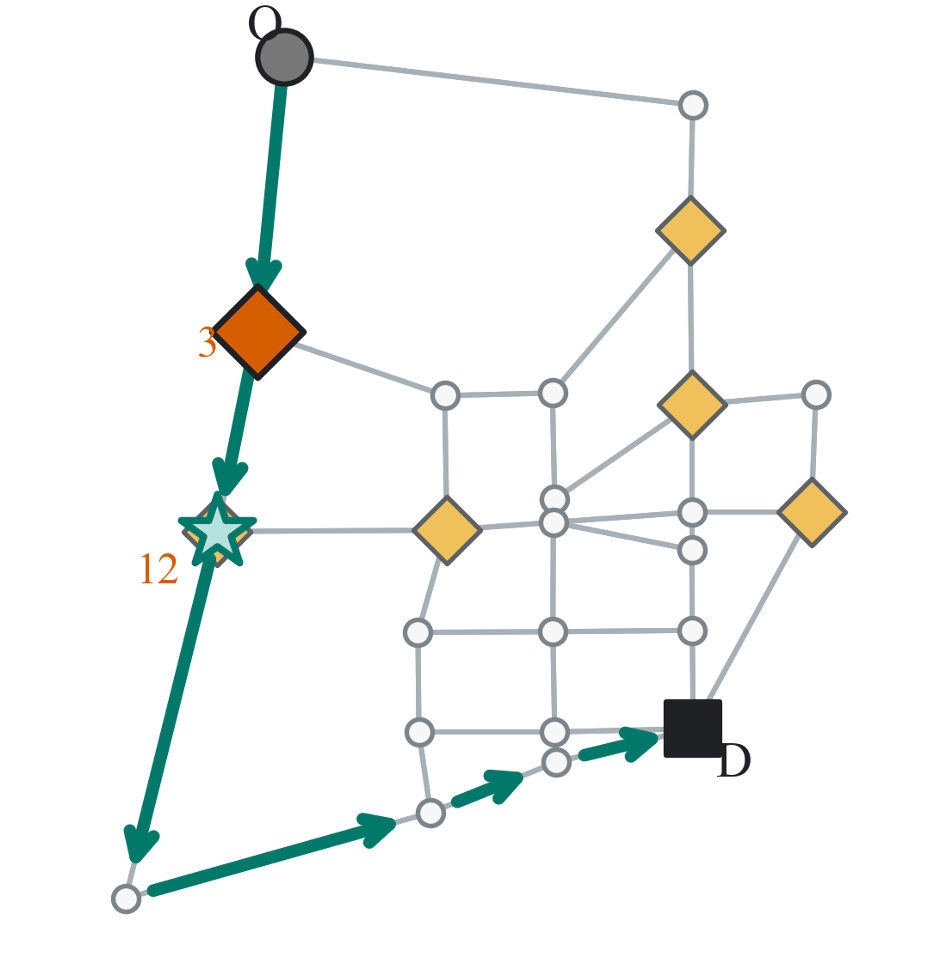}
\caption{V2G: 24km, 14.7 USD eq. Path:1-3-12-13-24-21-20; discharge at FCS 12, 8.4 kWh/trip.}
\label{fig:reroute_b}
\end{subfigure}

\caption{ Longer routes, lower cost: V2G-Driven detour behavior. Rerouting example for OD pair 3.}
\label{fig:reroute}
\end{figure}

\noindent
Figure~\ref{fig:reroute}
illustrates a representative rerouting outcome for OD pair 3. In the absence of V2G, the dominant path is 1-2-6-8-7-18-20, a 22 km route that includes no discharge stops, resulting in a relatively high total generalized cost. When V2G is available the equilibrium shifts to path 1-3-12-13-24-21-20, which, although longer in terms of time and distance, incorporates a discharge stop at FCS-12, yielding a lower total generalized path cost. This example demonstrates that V2G can alter route choices by improving overall trip economic, even when travel time and distance increase. 

\vspace{-0.1in}
\subsection{Overall System Cost}\label{sec:overall_cost}
\begin{figure}[!t]
\centering
\begin{subfigure}{1\linewidth}
\centering
\includegraphics[width=0.9\linewidth]{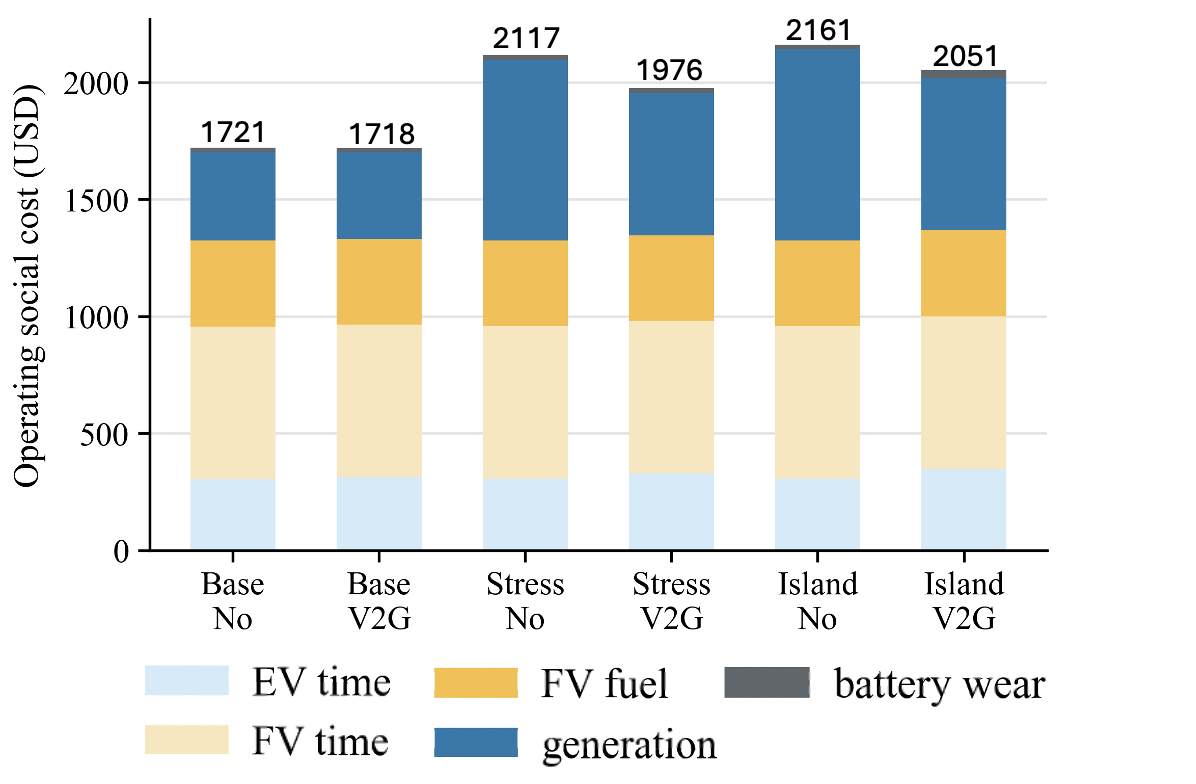}
\caption{Social-cost components excluding shortage}
\label{fig:social_cost_a}
\end{subfigure}
\hfill

\begin{subfigure}{1\linewidth}
\centering
\includegraphics[width=0.8\linewidth]{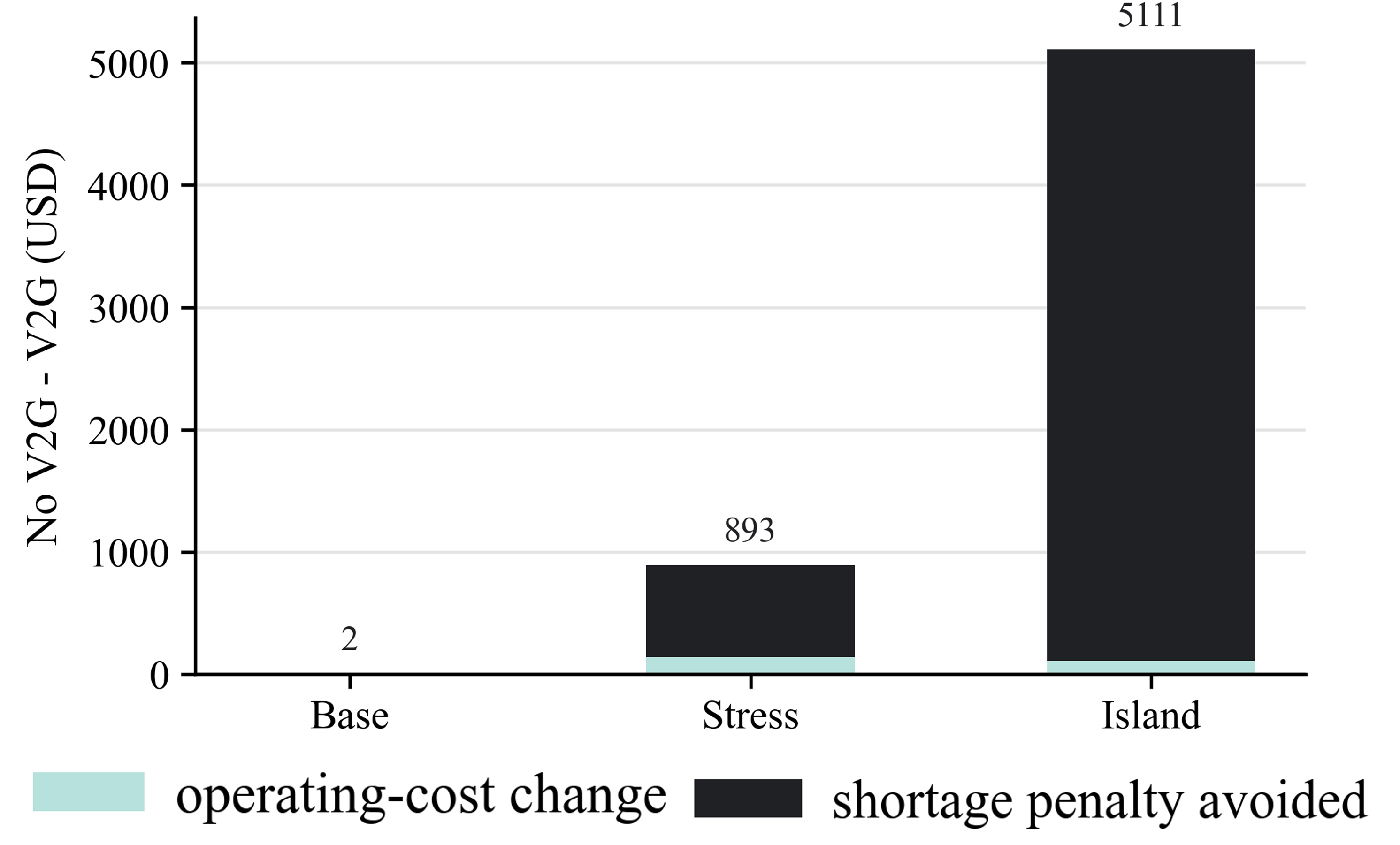}
\caption{Social-cost reduction}
\label{fig:social_cost_b}
\end{subfigure}

\caption{Analysis of total system social costs}
\label{fig:social_cost}
\end{figure}

\noindent To measure the aggregate effect of V2G across the integrated system, we evaluate the social cost.
Figure~\ref{fig:social_cost_a} presents the decomposition of operational social cost across scenarios. In the \textbf{Base} Case, enabling V2G yields only a marginal reduction in total social cost. Under the \textbf{Stress} and \textbf{Island} scenarios, however, V2G reduces total social cost by around 6\%, driven primarily by a marked decrease in generation costs.
As shown in Figure~\ref{fig:social_cost_b},V2G integration yields significant social-cost savings in high-stress environments, primarily through the mitigation of shortage penalties, confirming that the social benefit of V2G is strongly positive during grid emergencies.

\section{Conclusion}\label{sec:conclusion}

\noindent This paper developed a multi-player equilibrium framework to model the coupled interactions between transportation and power systems under V2G integration. By explicitly capturing the decision-making processes of CNO, EV users, DSO, and LSEs, the proposed model characterizes how system-wide outcomes emerge endogenously through market-clearing mechanisms and price signals. The resulting VI formulation admits a solution, provide the constraints of the players' decision problems and the market-clearing conditions are consistent. 
The numerical results demonstrate that V2G has minimal impact under normal operating conditions but becomes critically important under stressed and contingency scenarios. In particular, when the power system experiences high demand or network disruptions, V2G-enabled EVs provide localized energy support that mitigates scarcity, stabilizes DLMPs, and prevents load shedding. These effects are achieved through economically driven behavioral changes, where EV users respond to price incentives by adopting discharge-enabled routes and strategically interacting with FCSs.
From a system perspective, the integration of V2G significantly enhances resilience and reduces total social cost, primarily by lowering generation costs and eliminating shortage penalties. Importantly, 
the results highlight that user participation is not automatic but strongly dependent on economic incentives, emphasizing the importance of properly designed pricing mechanisms.
Overall, this work provides a rigorous analytical framework to evaluate V2G adoption under realistic behavioral and market conditions. 
Future research may extend this framework to other settings, such as the incorporation of stochastic renewable generation, heterogeneous user preferences, battery constraints, and location design of FCSs.

\bibliographystyle{IEEEtran}
\bibliography{references}

\end{document}